\documentclass[]{aa}  

\usepackage{graphicx}
\usepackage{txfonts}
\usepackage{amssymb}
\usepackage{multirow}
\usepackage{float}
\usepackage{gensymb}
\usepackage{xcolor}
\usepackage{rotating}
\usepackage[]{hyperref}
\hypersetup{backref=true, pagebackref=true, hyperindex=true, breaklinks=true,colorlinks=true,urlcolor=blue, linkcolor=blue, citecolor=blue,pagecolor=red, bookmarks=true, bookmarksopen=true}

\usepackage{amstext}

\begin{document}

   \title{Disks Around T Tauri Stars with SPHERE (DARTTS-S) II: \\ Twenty-one new polarimetric images of young stellar disks \thanks{Based on observations performed with VLT/SPHERE under program 098.C-0486(A).}}

   \subtitle{}

   \author{A.\,Garufi \inst{\ref{Firenze}}
   \and H.\,Avenhaus\inst{\ref{Klagenfurt}}
   \and S.\,P\'erez\inst{\ref{USantiago}}
   \and S.P.\,Quanz\inst{\ref{ETH}}
   \and R.G.\,van Holstein\inst{\ref{Leiden},\ref{ESO_Chile}}
   \and G.H.-M. Bertrang \inst{\ref{MPIA}} 
   \and \\ S.\,Casassus \inst{\ref{UChile}}
   \and L.\,Cieza\inst{\ref{Portales}}
   \and D.A.\,Principe\inst{\ref{Kavli}}
   \and G.\,van der Plas\inst{\ref{IPAG}}
   \and A.\,Zurlo\inst{\ref{Portales}} 
     }

\institute{INAF, Osservatorio Astrofisico di Arcetri, Largo Enrico Fermi 5, I-50125 Firenze, Italy. \label{Firenze}
  \email{agarufi@arcetri.astro.it}  
\and Lakeside Labs, Lakeside Park B04b, 9020 Klagenfurt, Austria \label{Klagenfurt}
\and Universidad de Santiago de Chile, Av. Libertador Bernardo O'Higgins 3363, Estaci\'on Central, Santiago, Chile \label{USantiago}
 \and Institute for Particle Physics and Astrophysics, ETH Zurich, Wolfgang-Pauli-Strasse 27, 8093 Zurich, Switzerland \label{ETH}
   \and Leiden Observatory, Leiden University, PO Box 9513, 2300 RA Leiden, The Netherlands \label{Leiden}
     \and European Southern Observatory, Alonso de C\'{o}rdova 3107, Casilla 19001, Vitacura, Santiago, Chile \label{ESO_Chile} 
        \and Max Planck Institute for Astronomy, K\"{o}nigstuhl 17, 69117 Heidelberg, Germany \label{MPIA} 
\and Departamento de Astronom\'{i}a, Universidad de Chile, Casilla 36-D Santiago, Chile   \label{UChile}
     \and Facultad de Ingenier\'ia y Ciencias, N\'ucleo de Astronom\'ia, Universidad Diego Portales, Av. Ejercito 441, Santiago, Chile \label{Portales} 
     \and Kavli Institute for Astrophysics and Space Research, Massachusetts Institute of Technology, Cambridge, MA 02139, USA \label{Kavli}
       \and Univ. Grenoble Alpes, CNRS, IPAG, F-38000 Grenoble, France \label{IPAG}
             }

   \date{Received -; accepted -}

 
\abstract{Near-IR polarimetric images of protoplanetary disks enable us to characterize substructures that might be due to the interaction with (forming) planets. The available census is strongly biased toward massive disks around old stars, however.}{The DARTTS program aims at alleviating this bias by imaging a large number of T Tauri stars with diverse properties.}{DARTTS-S employs VLT/SPHERE to image the polarized scattered light from disks. In parallel, DARTTS-A provides ALMA images of the same targets for a comparison of different dust components. In this work, we present new SPHERE images of 21 circumstellar disks, which is the largest sample released to date. We also recalculated some relevant stellar and disk properties following \textit{Gaia} DR2.}{The targets of this work are significantly younger than those published thus far with polarimetric near-IR (NIR) imaging. Scattered light is unambiguously resolved in 11 targets, and some polarized unresolved signal is detected in 3 additional sources. Some disk substructures are detected. However, the paucity of spirals and shadows from this sample {reinforces the trend according to which} these NIR features are {associated with Herbig stars, either because they are older or more massive}. Furthermore, disk rings that are apparent in ALMA observations of some targets do not appear to have corresponding detections with SPHERE. Inner cavities {larger than $\sim$15 au} are also absent from our images, even though they are expected from the spectral energy distribution. On the other hand, 3 objects show extended filaments at larger scale that are indicative of strong interaction with the surrounding medium. All but one of the undetected disks are best explained by their limited size {($\lesssim$20 au),} and the high occurrence of stellar companions in these sources suggests an important role in limiting the disk size. One undetected disk is massive and very large at millimeter wavelengths, implying that it is self-shadowed in the NIR.}{This work paves the way toward a more complete and less biased sample of scattered-light observations, which is required to interpret how disk features evolve throughout the disk lifetime.}


\authorrunning{Garufi et al.}

\titlerunning{Disks around TTSs with SPHERE}

   \maketitle
%

\section{Introduction} \label{Introduction}
Planets form in protoplanetary disks, and as a result, leave an observable imprint on them. This basic notion has motivated the community to focus on high-resolution imaging of circumstellar disks. Resolving disk substructures that may form as a result of gravitational interaction with (forming) planets for nearby star-forming regions requires a spatial resolution of about $\lesssim$0.1\arcsec. For the purpose of imaging disks, these resolutions are achieved by optical and near-IR (NIR) 8m telescopes and by the Atacama Large Millimeter Array (ALMA). In the visible and NIR, observations of protoplanetary disks require differential techniques to remove the dominating stellar flux. The most successful of these techniques employs a polarimeter to separate the stellar light from the (polarized) light that is scattered off the disk surface.

The NIR and ALMA images of planet-forming disks are complementary. In the NIR, observations are mostly sensitive to small dust grains, on the order of a $\text{micron}$ in size, while ALMA can trace both larger grains, on the order of a millimeter, and {polar molecules that provide rotational lines} in the gaseous phase. Furthermore, the NIR only probes the disk surface because disks are optically thick at these wavelengths, while ALMA can access deeper layers depending on the exact observing frequency and disk opacity. A few hundred planet-forming disks have thus far been imaged in the NIR or in the (sub-)millimeter, although only approximately 100 of them are high-resolution images. Disks imaged at both wavelengths with high resolution show strong morphological differences as well as some striking analogies. Rings and annular gaps have been recurrently imaged with both techniques \citep[e.g.,][]{Fedele2017, vanBoekel2017}. This is also true for large inner cavities \citep[e.g.,][]{Avenhaus2014a, vanderPlas2016} {and spiral arms \citep[e.g.,][]{Muto2012, Perez2016}. However, the latter type of structure is typically only observed in the millimeter around T Tauri stars (TTSs) \citep{Kurtovic2018} and in the NIR around Herbig AeBe and F stars \citep{Benisty2015}.} On the other hand, azimuthal asymmetries \citep[e.g.,][]{Casassus2013} and shadows are peculiarities of ALMA and NIR images, respectively. In particular, shadows at the disk surface \citep[e.g.,][]{Mayama2012, Marino2015, Stolker2016a} are seen as radially extended dark lanes and are the result of the inner disk geometry.

\begin{table*}
\centering
\caption{Main properties of the sample. Columns are the reference number in this work, the target name, the star-forming region, the distance $d$, the spectral type, the projected separation of bright visual companions, the stellar luminosity $L_*$, the mass $M_*$, the age $t$, the polarized-to-star light contrast $\delta_{\rm pol/star}$ (see Appendix \ref{Appendix_contrast}), the millimeter flux $F_{\rm mm}$ (at wavelengths given by the apex numbers), and the reference for $T_{\rm eff}$ (numbers) and $F_{\rm mm}$ (letters). Only bright companions (>1\% in flux of the primary) are listed here (a complete census can be found in Appendix \ref{Appendix_companion}). Stellar properties refer to the primary object (see Sect.\,\ref{Stellar_properties}). The full name of each target is shown in Table \ref{Observing_setup}.} 
\label{Sample_properties}
\begin{tabular}{ccccccccccccc}
\hline
Ref. & Target & Region & $d$ & SpT & ** & $L_*$ & $M_*$ & $t$ & $\delta_{\rm pol/star}$ & $F_{\rm mm}$ $^{\lambda}$ & Ref. \\
n. & name & & (pc) & & (\arcsec) & ($\rm L_{\odot}$) & ($\rm M_{\odot}$) & (Myr) & ($\times 10^{-3}$) & (mJy $^{\rm mm}$) & \\
\hline
1 & WW Cha & Chamaeleon & 191.9 & K5 & - & 11.1$\pm$2.0 & 1.0$\pm$0.2 & 0.2$^{+0.1}_{-0.1}$ & 3.9$\pm$0.6 & 1363 $^{0.87}$ &  1, a \\
\noalign{\smallskip}
2 & Sz 45 & Chamaeleon & 188.3 & M0 & - & 0.43$\pm$0.02 & 0.6$\pm$0.1 & 2.8$^{+0.6}_{-1.3}$ & 1.3$\pm$0.6 & 47.8 $^{1.3}$ & 2, b \\
\noalign{\smallskip}
3 & IK Lup & Lupus & 155.3 & K7 & 6.3 & 0.89$\pm$0.04 & 0.6$\pm$0.1 & 1.1$^{+1.0}_{-0.4}$ & 0.7$\pm$0.3 & 29.9 $^{1.3}$ & 3, c \\
\noalign{\smallskip}
4 & HT Lup & Lupus & 154.1 & K3 & 0.16 & 6.37$\pm$0.44 & 1.3$\pm$0.2 & 0.5$^{+0.4}_{-0.2}$ & 1.7$\pm$0.3 & 77 $^{1.3}$ & 4, d \\
\noalign{\smallskip}
5 & J1603-2031 & Upper Sco & 142.6 & K5 & - & 0.68$\pm$0.03 & 0.9$\pm$0.1 & 4.0$^{+4.0}_{-1.6}$ & <0.3 & 4.30 $^{0.88}$ & 3, e \\
\noalign{\smallskip}
6 & J1606-1928 & Upper Sco & - & M0 & 0.57 & 0.31$\pm$0.15 & 0.7$\pm$0.2 & 4.5$^{+11.4}_{-3.2}$ & <0.3 & 4.08 $^{0.88}$ & 3, e \\
\noalign{\smallskip}
7 & J1606-1908 & Upper Sco & 144.3 & K6 & 0.21 & 0.35$\pm$0.02 & 0.8$\pm$0.1 & 7.1$^{+8.2}_{-2.9}$ & <0.3 & 0.84 $^{0.88}$ &  5, e \\
\noalign{\smallskip}
8 & HK Lup & Lupus & 156.2 & M0 & - & 0.84$\pm$0.04 & 0.5$\pm$0.1 & 0.8$^{+0.5}_{-0.3}$ & <0.3 & 103.3 $^{1.3}$ & 6, c \\
\noalign{\smallskip}
9 & V1094 Sco & Lupus & 153.6 & K6 & - & 0.67$\pm$0.03 & 0.7$\pm$0.2 & 2.5$^{+2.0}_{-1.0}$ & 10.2$\pm$1.2 & 180 $^{1.3}$ & 7, f \\
\noalign{\smallskip}
10 & J1609-1908 & Upper Sco & 137.6 & M2 & - & 0.29$\pm$0.02 & 0.6$\pm$0.1 & 3.1$^{+2.4}_{-1.8}$ & 2.1$\pm$0.7 & 47.28 $^{0.88}$ & 8, e \\
\noalign{\smallskip}
11 & J1610-1904 & Upper Sco & - & M4 & 0.26 & 0.25$\pm$0.03 & 0.4$\pm$0.2 & 2.2$^{+3.0}_{-1.5}$ & <0.3 & 0.66 $^{0.88}$ & 5, e \\
\noalign{\smallskip}
12 & J1611-1757 & Upper Sco & 136.4 & M2 & - & 0.29$\pm$0.03 & 0.6$\pm$0.1 & 3.1$^{+2.4}_{-1.8}$ & <0.2 & <0.18 $^{0.88}$ & 5, e \\
\noalign{\smallskip}
13 & J1614-2305 & Upper Sco & - & K2 & 0.35 & 2.72$\pm$0.18 & 1.6$\pm$0.2 & 2.5$^{+1.6}_{-1.2}$ & <0.2 & 4.77 $^{0.88}$ & 4, e \\
\noalign{\smallskip}
14 & J1614-1906 & Upper Sco & 143.1 & M0 & - & 0.69$\pm$0.32 & 0.5$\pm$0.1 & 1.0$^{+2.4}_{-0.6}$ & 1.1$\pm$0.7 & 40.69 $^{0.88}$ & 5, e \\
\noalign{\smallskip}
15 & VV Sco & Upper Sco & 139.7 & K5 & 1.87 & 0.63$\pm$0.03 & 0.9$\pm$0.2 & 4.4$^{+3.6}_{-1.4}$ & <0.2 & 11.75 $^{0.88}$ & 4, e \\
\noalign{\smallskip}
16 & J1615-1921 & Upper Sco & 131.8 & K5 & - & 0.79$\pm$0.32 & 0.9$\pm$0.1 & 2.8$^{+7.1}_{-1.7}$ & 5.6$\pm$1.0 & 23.57 $^{0.88}$ & 4, e \\
\noalign{\smallskip}
17 & DoAr 16 & Ophiuchus & 137.4 & K4 & 0.81 & 2.26$\pm$0.26 & 1.1$\pm$0.2 & 1.0$^{+1.1}_{-0.5}$ & <0.3 & 47 $^{0.85}$ & 9, g \\
\noalign{\smallskip}
18 & SR4 & Ophiuchus & 134.6 & K0 & - & 3.46$\pm$0.53 & 1.7$\pm$0.1 & 2.8$^{+2.5}_{-1.4}$ & 1.1$\pm$0.2 & 70.4 $^{1.3}$ & 10, h \\
\noalign{\smallskip}
19 & DoAr 21 & Ophiuchus & 134.2 & K0 & - & 12.6$\pm$2.4 & 2.4$\pm$0.3 & 0.7$^{+0.6}_{-0.4}$ & <0.2 & <14 $^{0.85}$ & 11, h \\ 
\noalign{\smallskip}
20 & DoAr 25 & Ophiuchus & 138.5 & K5 & - & 1.67$\pm$0.19 & 0.8$\pm$0.2 & 0.9$^{+1.1}_{-0.3}$ & 8.4$\pm$0.8 & 239.0 $^{1.3}$ & 12, h \\
\noalign{\smallskip}
21 & SR9 & Ophiuchus & 130.4 & K5 & 0.66 & 1.72$\pm$0.10 & 0.8$\pm$0.1 & 0.9$^{+0.7}_{-0.3}$ & <0.2 & 6.3 $^{1.3}$ & 4, h \\
\noalign{\smallskip}
\hline
\end{tabular} 
\tablefoot{1 \citet{Luhman2007}, 2 \citet{Luhman2004}, 3 \citet{Koehler2000}, 4 \citet{Torres2006}, 5 \citet{Bouy2009}, 6 \citet{Herczeg2014}, 7 \citet{Cieza2007}, 8 \citet{Ansdell2016}, 9 \citet{Maheswar2003}, 10 \citet{Herbig1977}, 11 \citet{Bouvier1992}, 12 \citet{Ricci2010}; a \citet{Pascucci2016}, b \citet{Henning1993}, c \citet{Ansdell2018}, d \citet{Andrews2018}, e \citet{Barenfeld2016}, f \citet{vanTerwisga2018}, g \citet{Andrews2007}, h \citet{Cieza2019}.}
\end{table*}

ALMA surveys of star-forming regions \citep[e.g.,][]{Ansdell2016, Pascucci2016, Barenfeld2016, Long2018, Cieza2019} have exploited the possibility of obtaining snapshot exposures (of a few minutes) and have thus provided a quasi-complete census of protoplanetary disks {at $0.2\arcsec-0.7\arcsec$ resolution}. However, only a few campaigns such as the Disk Substructures at High Angular Resolution Project \citep[DSHARP,][]{Andrews2018} or individual studies have achieved very high angular resolutions ($\sim0.05\arcsec$). On the other hand, spatial resolutions are naturally high in the NIR, but snapshot exposures are not possible, and the time required to yield useful images is longer (tens of minutes to a few hours). This has led to mostly individual works focusing on one or a few specific sources \citep[e.g.,][]{Hashimoto2011, Quanz2013b, Benisty2015, Rapson2015, Bertrang2018}. As a result, the literature sample of disks imaged in the NIR is highly biased toward massive and large disks around bright stars \citep{Garufi2018}.

The Disks Around TTSs (DARTTS) program first presented by \citet{Avenhaus2018} has the objective to alleviate these biases by observing a relatively large number of TTSs with both the Spectro-Polarimeter High-contrast Exoplanet REsearch \citep[SPHERE,][]{Beuzit2019} and ALMA. The two parallel projects are called DARTTS-S (P.I.:H.\,Avenhaus) and DARTTS-A (P.I.:S.\,Perez). The first paper of the program \citep{Avenhaus2018} provided the first release of SPHERE images obtained for 8 millimeter-bright disks. In this work, we present the second release of the SPHERE dataset, including 21 new sources with diverse properties. To date, this sample represents the largest sample of NIR polarimetric images of protoplanetary disks ever released in a single work. Because the entire sample of disks available in the literature contains approximately 60 targets \citep[see][]{Garufi2018}, these new images correspond to roughly one-third of what has been published over the past decade. Therefore this study offers a significant boost to the demographical characterization of protoplanetary disks in scattered light. 

The paper is organized as follows. In Sect.\,\ref{Stellar_properties} we present the sample and the recalculation of all properties following \textit{Gaia} DR2 \citep{Gaia2018}. In Sect.\,\ref{Observations} we describe the observing setup and data reduction. The main results from the new SPHERE images are presented in Sect.\,\ref{Results} and discussed in Sect.\,\ref{Discussion}. We summarize our conclusions in Sect.\,\ref{Conclusions}.     
   
\section{Sample and stellar properties} \label{Stellar_properties}
The sample studied in this work consists of 21 TTSs from four different star-forming regions (Chamaeleon, Lupus, Upper Scorpius, and Ophiuchus), spanning spectral types K0 to M4 (see Table \ref{Sample_properties}). All these stars are known to host a circumstellar disk that has been imaged at (sub-)millimeter wavelengths (see Sect.\,\ref{Millimeter_images}). Compared to the first DARTTS-S release, this sample covers a much larger interval of disk dust masses, spanning the fluxes between a few hundred mJy at 1.3 mm and a fraction of mJy at 0.88 mm (see Table \ref{Sample_properties}).   

Nine targets are wide stellar binary systems, as is clear from inspection of our SPHERE intensity images (see Sect.\,\ref{Results}). To our knowledge, three of these stars (J1606-1928, J1606-1908, and J1610-1904) were not known to have a companion. For the other sources, the detected companions are present in the literature \citep{Ratzka2005, Torres2006, Metchev2009}. One of these targets is a triple system \citep[HT Lup,][]{Ghez1997}. The projected physical separations of the companions span from 25 au to nearly 1000 au, although 70\% of them are closer than 85 au. In addition to these nine wide systems, two of our sources are probable spectroscopic binaries \citep[DoAr 21 and WW Cha,][]{Loinard2008, Anthonioz2015}. Astrometric and photometric properties of all companions are described in Appendix \ref{Appendix_companion}.


To ensure homogeneity and to include the newest distance estimate from \textit{Gaia} DR2 \citep{Gaia2018}, we recalculated all stellar properties consistently with the larger sample by \citet{Garufi2018}. We collected the complete spectral energy distribution (SED) of each source through \textit{Vizier} and calculated the stellar luminosity through a PHOENIX model of the stellar photosphere \citep{Hauschildt1999} with effective temperature $T_{\rm eff}$ and extinction $A_{\rm V}$ taken from the literature and using the dereddened $V$ magnitude and the \textit{Gaia} DR2 distance $d$ as benchmarks to scale the model. Three sources, all in Upper Sco, have no available \textit{Gaia} distance. Interestingly, these three have a stellar companion between 0.25\arcsec\ and 0.6\arcsec \footnote{J1606-1928, J1610-1904, and J1614-2305 are flagged as \textsc{dup}=1 in the \textit{Gaia} catalog, indicating that a possible duplicate source was removed before publication. The duplicated source could also have been real companions, imperfectly cross-matched additional transits of the same source, artifacts, or a mixture of various effects. In the future \textit{Gaia} releases, a set of separate pipelines will deal with non-single sources, and the data quality is expected to improve in cases like these.} For these stars, we assumed \mbox{$d$=140 pc}, as this is the averaged distance of the other Upper Sco sources of our sample.

The construction of the SED allowed us to calculate the NIR and far-IR (FIR) excess of each source as in \citet{Garufi2017}. We also constrained the stellar age and mass through the PARSEC pre-main-sequence (PMS) tracks \citep{Bressan2012}. The values obtained, shown in Table \ref{Sample_properties}, are in reasonable agreement with those obtained with other PMS tracks \citep[Baraffe and \textsc{MIST},][]{Baraffe2015, Choi2016}. In case of close stellar systems, the luminosity of the primary star was extracted through the flux ratios calculated from our data (see Appendix \ref{Appendix_companion}). Finally, the dust disk masses were estimated through the millimeter fluxes under standard assumptions, as in \citet{Garufi2018}. These estimates should be regarded with caution given our poor knowledge of dust opacity and disk temperature \citep[see, e.g.,][]{Birnstiel2018} and the possibility that the inner disk regions are optically thick at millimeter wavelengths. 

As is clear from Table \ref{Sample_properties}, the targets studied in this work cover a relatively large interval of stellar masses (between 0.4 $\rm M_{\odot}$ and 2.5 $\rm M_{\odot}$), although only five of them are super-solar mass stars. Another peculiarity of the sample is, in the current framework of stellar tracks, the young age. An average 2.3$^{+3.9}_{-1.4}$ Myr is found. This value should be compared with the 6.5 Myr of the complete sample of circumstellar disks published thus far in the NIR \citep[see][]{Garufi2018}. These aspects are analyzed in detail in Sect.\,\ref{Discussion_sample}.

\section{Observations and data reduction} \label{Observations}
All our observations were carried out with the SPHERE Infra-Red Dual Imaging and Spectrograph \citep[IRDIS,][]{Dohlen2008} in dual polarization imaging mode \citep[DPI,][]{Langlois2014, deBoer2019, vanHolstein2019}. This technique allows an efficient suppression of the stellar (only marginally polarized) light while keeping the scattered (highly polarized) light from the circumstellar disk relatively intact. The observations were conducted between 7-13 March 2017 in the H band. We employed a coronagraph with a diameter of 185 mas \citep{Carbillet2011} to mask the central bright star while allowing for long exposure times (typically 96 s, with shorter times in a few cases). The total integration times varied from approximately 25 to 38 minutes (see Table \ref{Observing_setup} for details). In addition to the main science frames, we obtained center and flux complementary frames at various stages of the observation. These are needed to accurately determine the stellar position behind the coronagraph and its flux, respectively. 

The scientific products of this type of observations are a set of four linear polarization components, called $Q^+$, $Q^-$, $U^+$,  and $U^-$, that are obtained by subtracting the two beams with orthogonal polarization states recorded simultaneously on the detector and tuning their polarization direction with a half-wave plate with positions of 0$\degree$, 45$\degree$, 22.5$\degree$, and 67.5$\degree$, respectively. In this work, we employed the IRDAP\footnote{\url{irdap.readthedocs.io}} (IRDIS Data reduction for Accurate Polarimetry) pipeline, version 0.3.0 \citep{vanHolstein2019}. The Stokes $Q$, $U$, and $I$ images were obtained by computing the double differences from $Q^+$ and $Q^-$, from $U^+$ and $U^-$, and the double sum, respectively. IRDAP can differentiate the instrumental and stellar polarization. After computing the $Q$ and $U$ images, the pipeline uses the Mueller matrix model to subtract the instrumental polarization of each image and least squares to correct for the cross talk. The images thus created may still contain some stellar polarization that is constrained by measuring the flux in the $Q$ and $U$ images around regions that should be virtually devoid of polarized signal from the disk. The correction for the stellar polarization may or may not be implemented, depending on the scope of the analysis. Any unresolved material close to the star may generate a halo of polarized light in the images. This affects the outer regions, but also carries information on the inner regions. In this work, we make use of images that corrected (Sect.\,\ref{Overview}) and images that not corrected (Sect.\,\ref{Stellar_polarization}) for the stellar polarization.

\begin{figure*}
  \centering
 \includegraphics[width=17cm]{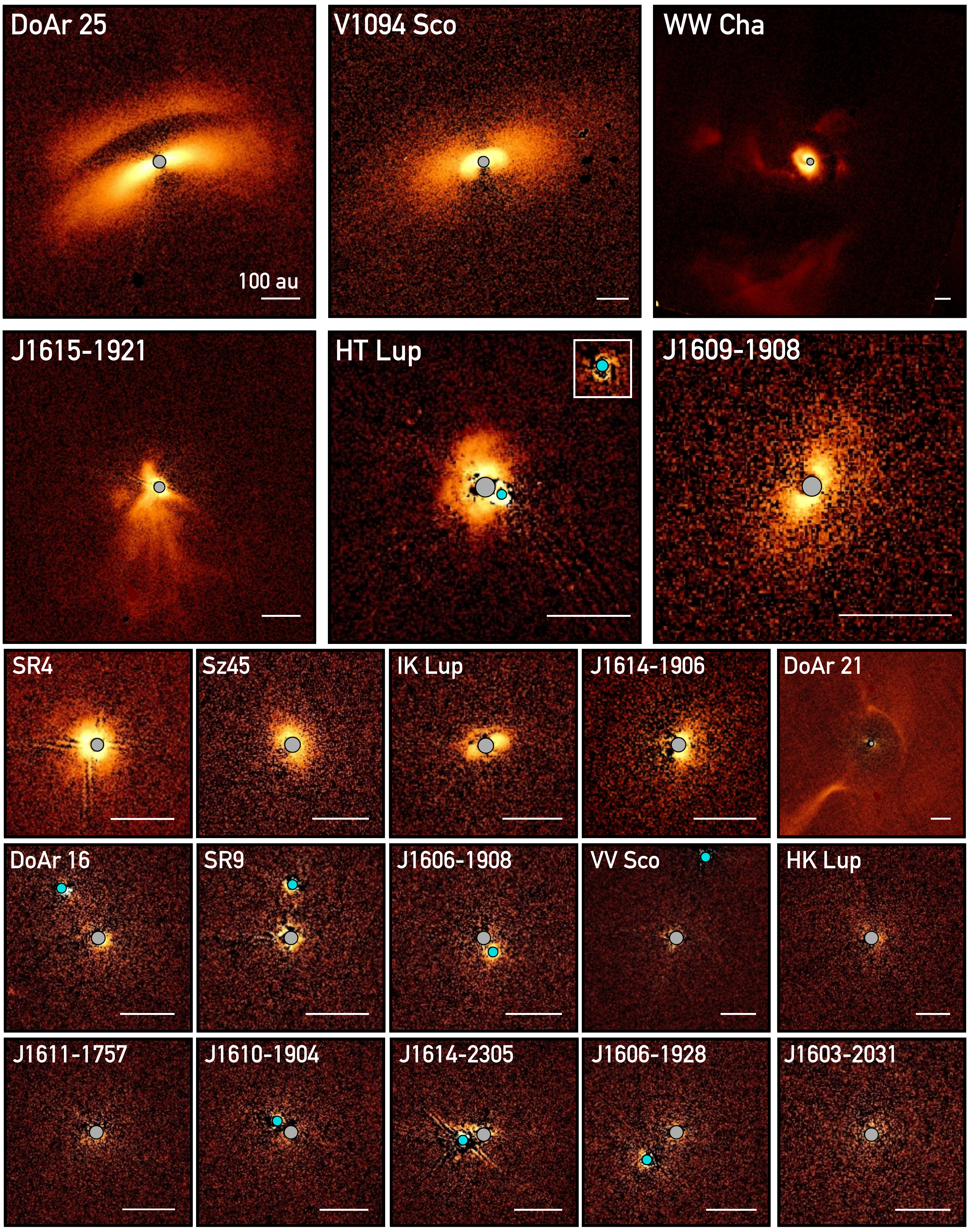} 
     \caption{Overview of the sample. For each target, the $Q_\phi$ image in the H band is shown. The first two rows host disks with clearly visible substructures, the third row contains disks with no or marginal evidence of substructures, and the last two rows show non-detections. The horizontal bars indicate a physical scale of 100 au. The main star is behind the gray circle in the center, symbolizing the coronagraph. Stellar companions are indicated by cyan circles. Each image has a different logarithmic color scale, chosen to highlight all relevant features. The C component of HT Lup is shown in the inset image, although the star would lie outside of the box. North is up, and east is left. }
          \label{Imagery}
 \end{figure*}

The azimuthal Stokes parameters $Q_\phi$ and $U_\phi$ \citep[see][]{deBoer2019} are the final product of our data reduction. The $Q_\phi$ image traces the azimuthally polarized flux and is therefore expected to be very similar to the polarized intensity $P\equiv \sqrt{Q^2+U^2}$ in case of centrosymmetric single scattering from disks with low inclination \citep{Canovas2015}. On the other hand, $U_\phi$ traces the component 45$\degree$ inclined with respect to the azimuthal direction and to first order contains only noise. Each of our final images is also corrected for true north \citep{Maire2016}, and the pixels are assumed to have a scale of 12.25 mas. 

\begin{figure*}
  \centering
 \includegraphics[width=17cm]{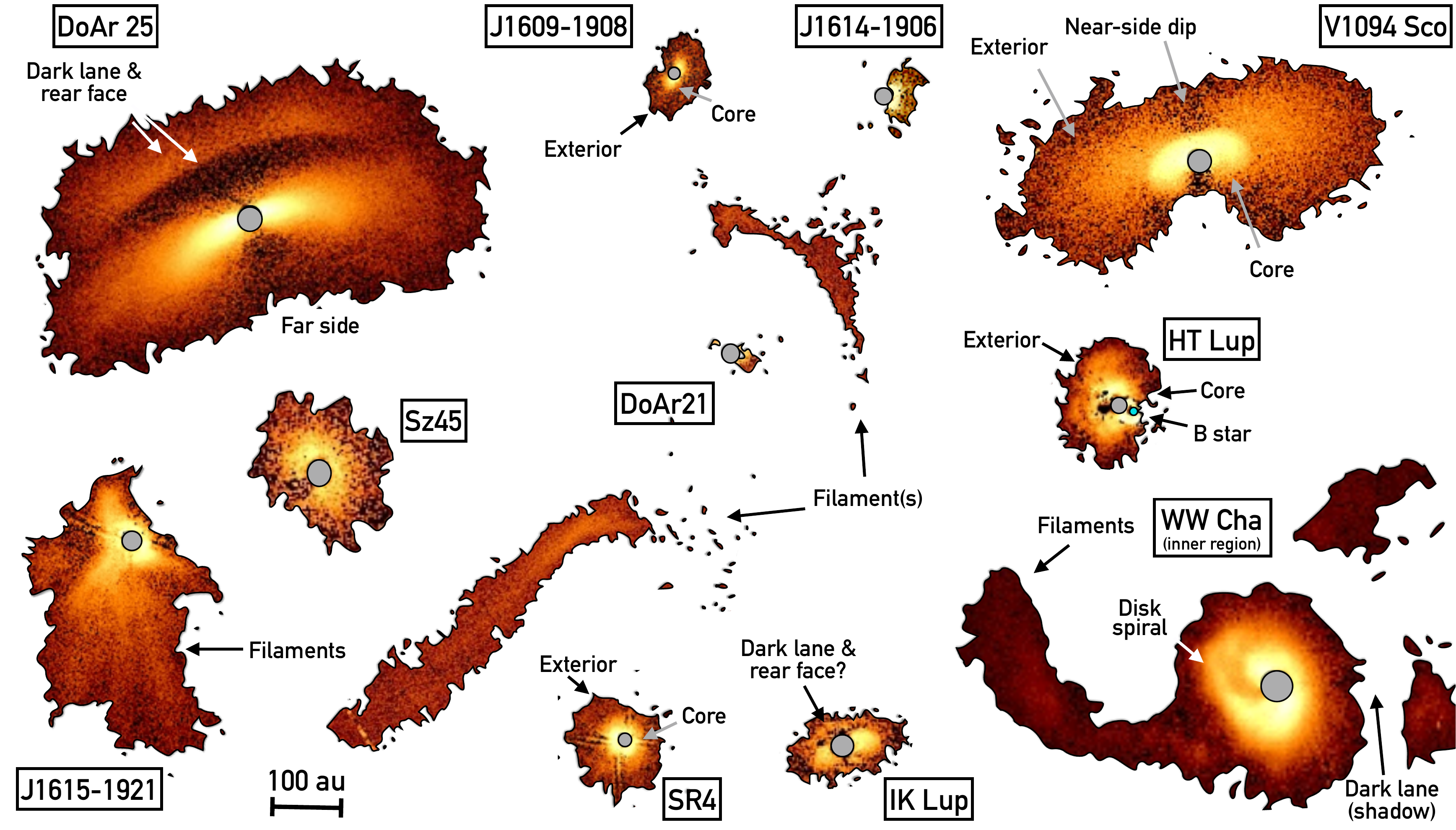} 
     \caption{Labeled version of Fig.\,\ref{Imagery}. The disks have the same spatial scale for comparison. Only detections are shown.}
          \label{Scaled_imagery}
 \end{figure*}

\section{Results} \label{Results}

\subsection{Overview} \label{Overview}
All the polarized images of this work are shown in Fig.\,\ref{Imagery}. Of the 21 targeted objects, 11 show unambiguous evidence of scattered light from their $Q_\phi$ image. Disk substructures are clearly visible in at least 6 objects (mainly the first two rows of Fig.\,\ref{Imagery}). In 3 objects (WW Cha, J1615-1921, and DoAr 21), extended filaments from the surrounding medium are evident. The polarized light contrast measured with respect to the central star spans from 10$^{-2}$ to 7$\times10^{-4}$ (see Table \ref{Sample_properties} and Appendix \ref{Appendix_contrast}). The detected disks also span a wide range in size, where the outermost radius reaches 300 au and the smallest disk extends to only 50 au. This wide range of disk sizes is apparent in Fig.\,\ref{Scaled_imagery}.   

Close stellar companions (<2\arcsec\ from the central star) are evident in eight images. To our knowledge, three of these were not known before (see Sect.\,\ref{Stellar_properties}). With one notable exception (HT Lup), the presence of a companion implies a non-detection of scattered light around the primary. Interestingly, seven of the ten non-detections have a detected close companion.

\subsection{Individual sources} \label{Individual}
DoAr 25 is by far the most spectacular detection from this sample (see Fig.\,\ref{Imagery}). Its image resembles that of IM Lup from the first DARTTS paper \citep{Avenhaus2018}, although the disk is evidently more inclined. Strong signal is detected from the disk front face out to 2.3\arcsec, corresponding to $\sim$320 au, while a dark lane with no detectable flux separates it from the rear disk face (see also Fig.\,\ref{Scaled_imagery}). This lane has a maximum width of 0.2\arcsec\ along the disk minor axis. From simple geometrical considerations, the sum of this width and that of the rear side (0.35\arcsec) yields the disk opening angle at the last scattering separation (2.3\arcsec), from which a disk flaring of $\sim$0.24 can be constrained at $\sim$300 au. This value is very similar to what was obtained for IM Lup by \citet{Avenhaus2018}, reinforcing the analogy between these two objects. On the disk front face of DoAr 25, the near side (closer to the dark lane) is two orders of magnitude brighter than the far side. This discrepancy is most likely entirely due to the tendency of photons to be scattered by small angles \citep[an effect yielding a forward-peaking scattering phase function, see, e.g.,][]{Stolker2016b}. Except for the disk silhouette, no particular disk substructures are evident from our images. In particular, no cavity or annular gap or ring is detected down to 13 au (see Sect.\,\ref{Millimeter_images}).

V1094 Sco is the second most prominent disk of the sample, being detected out to 1.8\arcsec\ ($\sim$270 au). Unlike DoAr 25, no disk silhouette is visible. The disk seems to be composed of two regions. An inner region (called core in Fig.\,\ref{Scaled_imagery}) extending out to 0.45\arcsec\ ($\sim$70 au) is very bright in scattered light. Just beyond it, the flux is abruptly diminished but then declines only shallowly outward (exterior emission). This outer region seems to show a localized dip along the minor axis. However, this effect may also be due to an imperfect removal of the stellar polarization (see Sect.\,\ref{Stellar_polarization}). 

WW Cha and J1615-1921 both show a complex morphology with a bright inner component around the star and asymmetric filaments extending along several directions out to the detector edge. The inner signal around WW Cha is maximized at angles of 50$\degree$ and 230$\degree$, suggesting a relatively inclined disk with a position angle of 50$\degree$ and an apparent extension of roughly 0.7\arcsec\ (130 au). A bright spiral arm is visible to the east wrapping toward north (clockwise outward). A dark lane separates this inner component from the structures to the west. Unlike DoAr 25, this cannot be the disk silhouette because it extends much farther than the disk itself. Instead, it is more likely a shadow cast by the disk on these outer filaments (see Ginski et al. in prep.). Interestingly, all the filaments visible from our image wrap in a clockwise direction (from inside out), in analogy with the disk spiral. 

Our image of the triple system HT Lup only reveals polarized signal around the primary star. This consists of both a bright region symmetric around the north-south axis and an exterior emission (see also Fig.\,\ref{Scaled_imagery}) to the east. This flux is clearly asymmetric, being only detected to the east in a crescent-like shape. The flux around the B component (close to the primary) and the C component (in the inset box of Fig.\,\ref{Imagery}) perfectly resembles that seen in the intensity image and is therefore only an unremoved point spread function (PSF).

The disk of J1609-1908 appears as a smaller version of that of V1094 Sco, extending only out to 0.6\arcsec\ (85 au). SR4 also shows a rather bright component around the coronagraph and a fainter exterior emission visible out to 0.5\arcsec\ ($\sim$70 au) toward the SW. The images of Sz45 and IK Lup suggest mildly inclined disks ($i>45\degree$) with a position angle (P.A.) of 40$\degree$ and 115$\degree$ and an apparent extension of 0.4\arcsec\ (75 au) and 0.35\arcsec\ (55 au), respectively. In IK Lup, there seems to be a dark lane to the north and a bright half-moon shape that suggest a smaller version of the disk geometry seen in DoAr 25. Strong signal is detected around J1614-1906 to the western side only, but any interpretation of the disk geometry based on this signal is probably doubtful. Finally, DoAr 21 shows a most peculiar morphology. A circumstellar disk may be marginally detected just outside the coronagraph to the east and west. More importantly, a bright filament stands out to the SE and proceeds counterclockwise, mostly disappearing to the south and then reappearing to the west toward a bright knot to the north. 

The remaining ten sources do not show any significant signal. Any marginal flux recorded in the $Q_\phi$ images in the near surrounding of the coronagraph resembles that of the intensity image and cannot be interpreted as a robust detection. Nevertheless, flux from this very inner region may still be real emission from an unresolved source, as described in Sect.\,\ref{Stellar_polarization}. As commented in Sect.\,\ref{Overview}, the incidence of close stellar companions in the subsample of non-detections is very high (70\%). The apparent distance of these companions spans from 0.22\arcsec\ (30 au) to 1.94\arcsec\ (270 au), although the second most distant is only at 0.80\arcsec\ (110 au). These findings are discussed in Sect.\,\ref{Discussion_non-detections}.

\subsection{Comparison with ALMA images} \label{Millimeter_images}
All the targets of this work have been observed with moderate-to-high resolution ALMA (sub-)millimeter imaging. The Chamaeleon sources have been discussed by \citet{Pascucci2016}, the Lupus sources by \citet{Ansdell2018}, the Ophiuchus sources by \citet{Cieza2019}, and the Upper Sco sources by \citet{Barenfeld2017}. {Three} targets have also been observed within the DSHARP program \citep{Andrews2018}. All disks detected in scattered light are resolved in the \mbox{(sub-)millimeter}. Only two disks in the whole sample of this work are not detected (J1611-1757 and DoAr21), and six are unresolved or marginally resolved by ALMA (see below).

In Fig.\,\ref{ALMA_SPHERE} we show the direct comparison for the four disks with substructures from ALMA. The millimeter image of V1094 Sco is a well-known example of disks with rings and annular gaps \citep{vanTerwisga2018}. Interestingly, these radial variations are not obvious in the SPHERE image. The only strong analogy between the two images is the core emission. In the ALMA image, \citet{vanTerwisga2018} revealed an optically thick, bright central core. This region has a perfect counterpart in our image, suggesting an abrupt vertical discontinuity at the location where the disk becomes optically thin in the millimeter. 

Similarly to V1094 Sco, DoAr 25 shows disk rings and annular gaps \citep{Andrews2018} that are not detected in the SPHERE images. Instead, we cannot rule out the presence of an NIR counterpart to the disk gap seen by ALMA around SR4 because this lies only marginally outside of the SPHERE coronagraph. The millimeter outer ring of this disk is spatially consistent with the bright core emission from SPHERE, while the exterior emission is not detected by ALMA. This suggests another connection between the vertical extension of the disk and the physical properties at the midplane.  

The continuum emission around the primary star HT Lup A also corresponds to the bright inner region visible from SPHERE. The spiral structure detected from this region \citep{Andrews2018, Kurtovic2018} is not visible in our image. From the comparison, it is nonetheless clear that such a detection would be very challenging from our data given the proximity to the coronagraph. Analogously, the continuum emission around the B star is clearly too small ($\sim$3 mas) to be detected in our image because the flux from the star itself is relatively high. On the other hand, the $^{12}$CO emission extends to radii that are comparable to our exterior emission. The gas emission appears symmetric around the minor axis, however, while the east side from SPHERE is significantly brighter.

The coarser {resolution ($0.5\arcsec \times 0.7\arcsec$) of the} ALMA image of WW Cha \citep{Pascucci2016} discourages the direct comparison with the SPHERE image. It is nonetheless important to note that the extended filaments detected in scattered light do not have any millimeter counterpart. The spiral structure visible in the NIR was not visible from ALMA, although an asymmetry in the same direction seems to be present. Similarly to WW Cha, the millimeter continuum image of J1615-1921 only shows the inner emission from the disk \citep[fitted with an outer radius of only 10 au,][]{Barenfeld2017}. Nevertheless, Barenfeld and collaborators also showed the $^{12}$CO map of this object and revealed that the emission is much more extended (430 au) and is contaminated by the surrounding molecular cloud. An asymmetric elongation toward the south is visible in their image, and this morphology resembles our observed filaments. Finally, no millimeter emission is detected around DoAr 21 \citep{Cieza2019}, which as for WW Cha and J1615-1921 rules out the substantial presence of millimeter grains in the filaments seen in scattered light. 

\begin{figure}
  \centering
 \includegraphics[width=9cm]{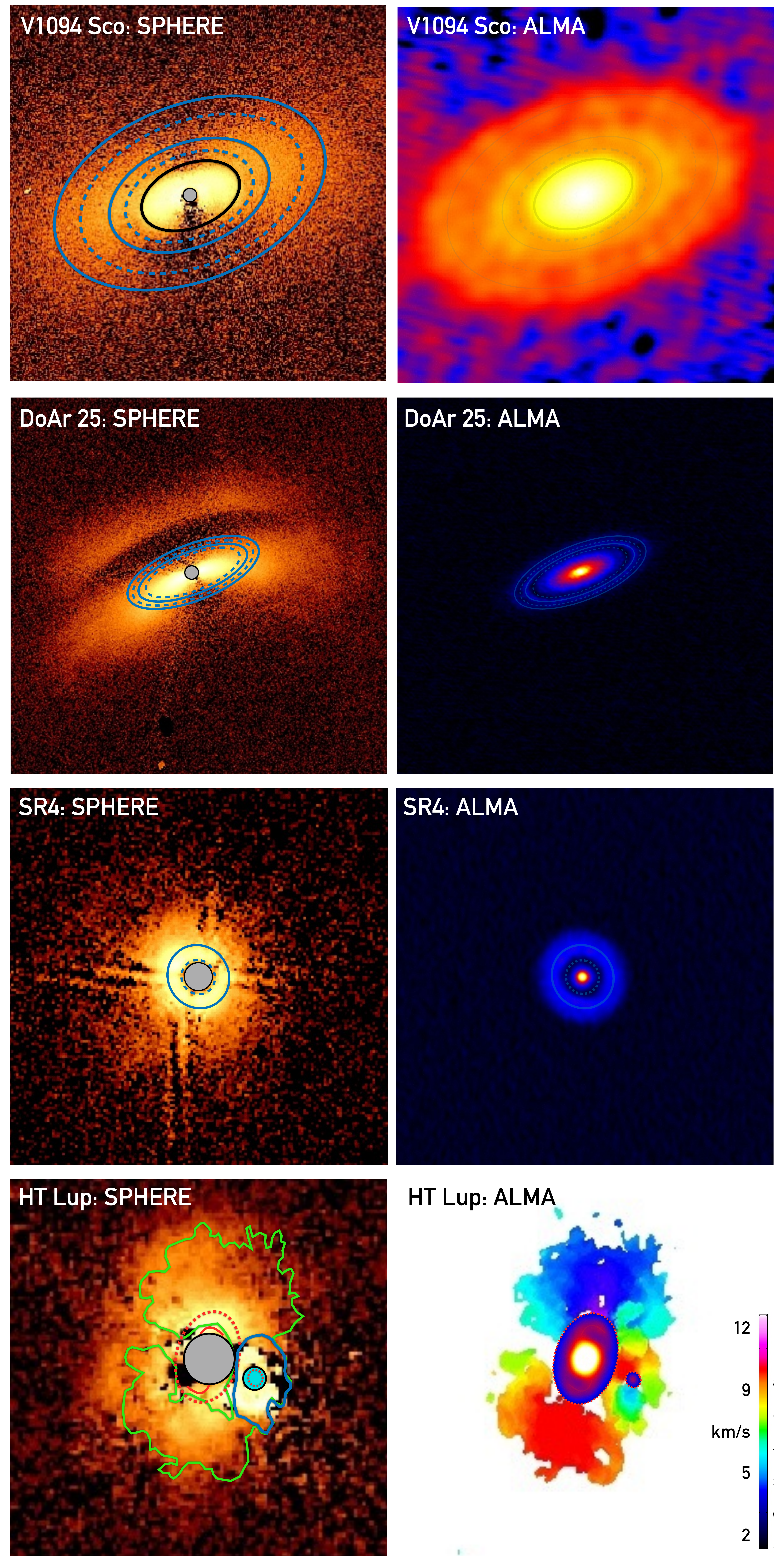} 
     \caption{Comparison with ALMA images in Band 6. \textit{First three rows:} SPHERE images of V1094 Sco, DoAr 25, and SR4 (left) compared with the continuum ALMA images (right). Rings and gaps from the ALMA image are plotted on top of the SPHERE image with full and dashed lines, respectively. The black line in V1094 Sco indicates the inner core. \textit{Bottom row:} SPHERE image of HT Lup (left) compared with the continuum (inset ellipses) and $^{12}$CO emission (main figure) from ALMA (right). The color wedge delineates the $^{12}$CO emission. The red, green, and blue lines in the SPHERE image represent the outer edges of the continuum emission, the total disk CO emission, and CO emission associated with the B star, respectively.}
               \label{ALMA_SPHERE}
 \end{figure}

\begin{figure}
  \centering
 \includegraphics[width=8.5cm]{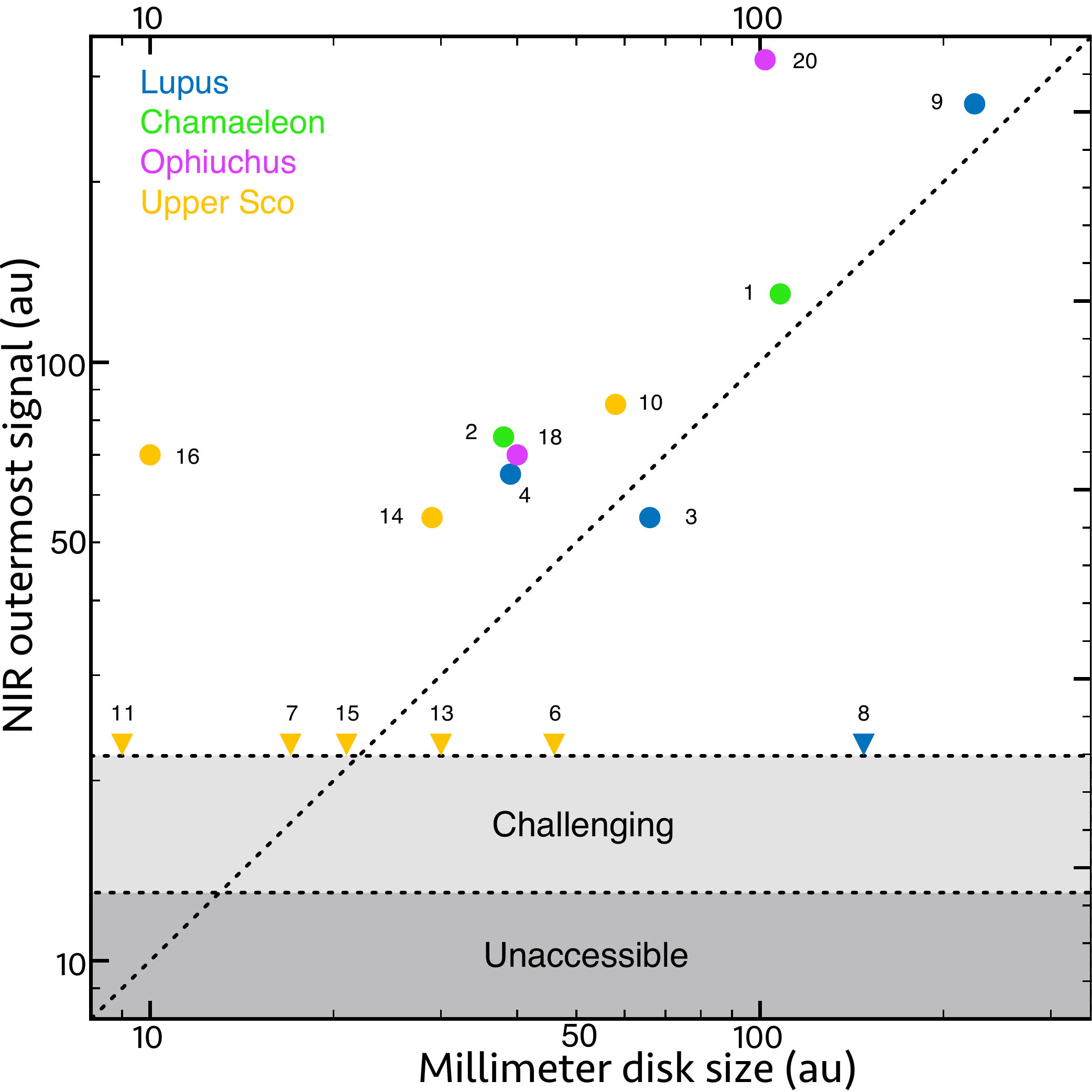} 
     \caption{Near-IR vs. millimeter disk sizes. Only targets with resolved millimeter observations are shown. Disks that are not detected in polarized light are indicated by the triangle. Their value along the y-axis is imposed as the separation from the central star where NIR detections become challenging (0.15\arcsec$\times$<$d$>). The diagonal line indicates the equality of the two axes. Target numbers refer to Table \ref{Sample_properties}.}
          \label{Outer_radii}
 \end{figure}

All the disk sizes constrained from ALMA continuum images {(see the aforementioned references)} can be compared with the apparent disk sizes revealed with SPHERE. {These are constrained by the separation in our images where the signal drops below 3$\sigma$.} The comparison is shown in Fig.\,\ref{Outer_radii}. Considering only detections, the disks in the NIR are on average $\sim$50\% larger than in the millimeter. Two important outliers are DoAr 25 (number 20) and J1615-1921 (number 16), showing a ratio of 300\% and 700\%, respectively. For the latter, this high ratio is probably explained by the fact that our images do not trace the actual disk but rather the filaments that reach the immediate surrounding of the star. More generally, the apparent trend between NIR and millimeter sizes suggests that {some of our non-detections are likely explained by a very small disk. On the other hand, this does not apply to J1606-1928 (number 6) and HK Lup (number 8) because their non-detection is clearly inconsistent with the observed trend} (see Sect.\,\ref{Discussion_non-detections}).

\subsection{Central polarization} \label{Stellar_polarization}
\begin{figure*}
  \centering
 \includegraphics[width=18.5cm]{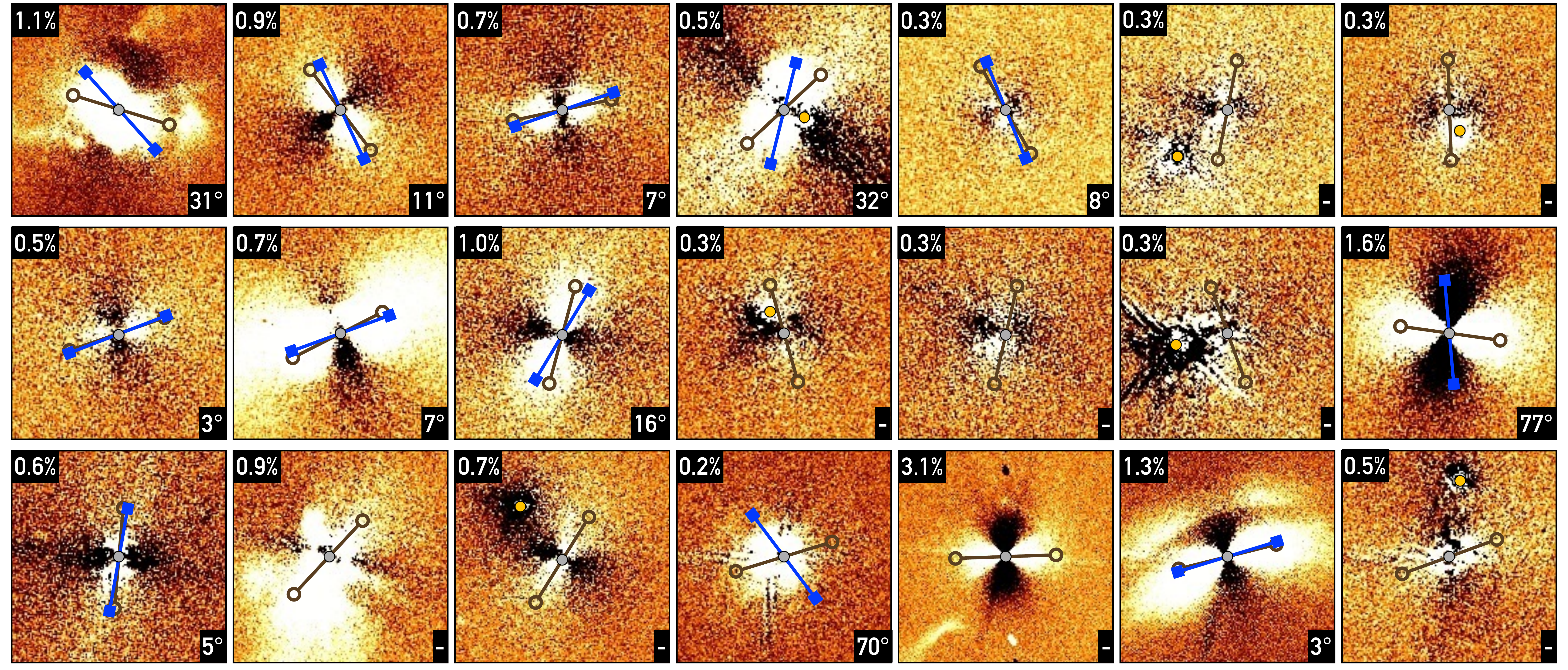} 
     \caption{Central polarization in our sample. For each target, the $Q_\phi$ image is corrected for the instrumental polarization but not corrected for the stellar polarization. The brown line with empty circles is perpendicular to the polarization direction, while the percentage in the top right corner indicates the degree of polarization. Where it is known, the position angle of the outer disk is indicated by the blue line with filled squares. The misalignment between the two orientations is expressed in the bottom right corner. Targets are sorted as in Table \ref{Sample_properties} with the three rows hosting numbers 1$-$7, 8$-$14, and 15$-$21.}
          \label{Stellar_pol}
 \end{figure*}

As we described in Sect.\,\ref{Observations}, the pipeline IRDAP allows us to separate the instrumental and stellar polarization and therefore to study the latter individually. All our targets show evidence of stellar polarization, and the degree of linear polarization spans from 0.2\% to 3.1\%, although the majority of targets (17/21) show less than 1.0\%. This polarization is visible in the $Q_\phi$ images before its subtraction (see Sect.\,\ref{Observations}), as shown in Fig.\,\ref{Stellar_pol}. It is clear that most of our sources show a butterfly pattern, indicating a central unresolved polarization source. Any spatially unresolved polarized signal is smeared out when it is convolved with the PSF and thus generates a characteristic butterfly pattern at larger scale of the $Q_\phi$ image. This polarization is particularly interesting in the scenario where it originates from an unresolved or marginally resolved inner disk \citep[see][]{vanHolstein2019}. If this is confirmed, the measured angle of linear polarization is perpendicular to the disk major axis because photons are more highly polarized for a scattering angle close to 90$\degree$ \citep[e.g.,][]{Murakawa2010}.

To test whether these patterns can be due to an unresolved inner disk, in Fig.\,\ref{Stellar_pol} we compare the angle of the stellar polarization (rotated by $90\degree$) with the position angle of the outer disk constrained from ALMA {(see references in Sect.\,\ref{Millimeter_images})}, when this is known with sufficient precision ($<\pm20\degree$). For 8 of the 12  targets that can be analyzed, the misalignment between these two angles is smaller than 16$\degree$. In turn, in 3 of these 8 targets (J1603-2031, HK Lup, and VV Sco) we do not resolve the disk (see Fig.\,\ref{Imagery}). Because the misalignment measured for these sources is as small as 7$\degree$, 3$\degree$, and 5$\degree$, we can argue with good confidence that the source of measured polarization is the unresolved inner disk. For the other 5 of these 8 targets, this alignment rules out dramatic changes in the disk orientation (disk warps) between a few au and some hundred au. The other 4 of the 12 targets that can be analyzed show a stronger misalignment, from 31$\degree$ to 77$\degree$. For 3 of them, the misalignment is most likely due to the interstellar material that polarizes the stellar beam as a result of the higher measured extinction $A_{\rm V}$ of 3.0, 4.0, and 4.8 (whereas the 8 aligned cases have lower $A_{\rm V}$). The remaining source, HT Lup, with a misalignment of 33$\degree$, is probably explained by the presence of the companion that carries its own disk close to the coronagraph (Fig.\,\ref{Imagery}). 

The results found for this relatively large sample therefore support the view for which in case of low interstellar extinction ($A_{\rm V}\ll3$), any measured stellar polarization indicates the presence of an unresolved (inner) disk and the angle of linear polarization is likely related to its geometry, although the disk is not resolved in the final $Q_\phi$ image \citep[see also][]{Keppler2018}. As a consequence, we may realistically also claim the detection of scattered light from 6 of the 9 targets (those with low $A_{\rm V}$) with no robust constraint on the disk geometry from ALMA. These are J1606-1908, J1606-1928, J1610-1904, J1611-1757, J1614-2305, and SR9.

\section{Discussion} \label{Discussion}
We focus much of the discussion on the demography of disks in scattered light following the release of this new dataset. This is motivated by the large size of our sample, which increases the total number available in the literature by more than one-third. Further analyses on individual sources are still certainly worthwhile, but we defer this effort to future dedicated works. 

\subsection{This sample in context} \label{Discussion_sample}
\begin{figure*}
  \centering
 \includegraphics[width=9cm]{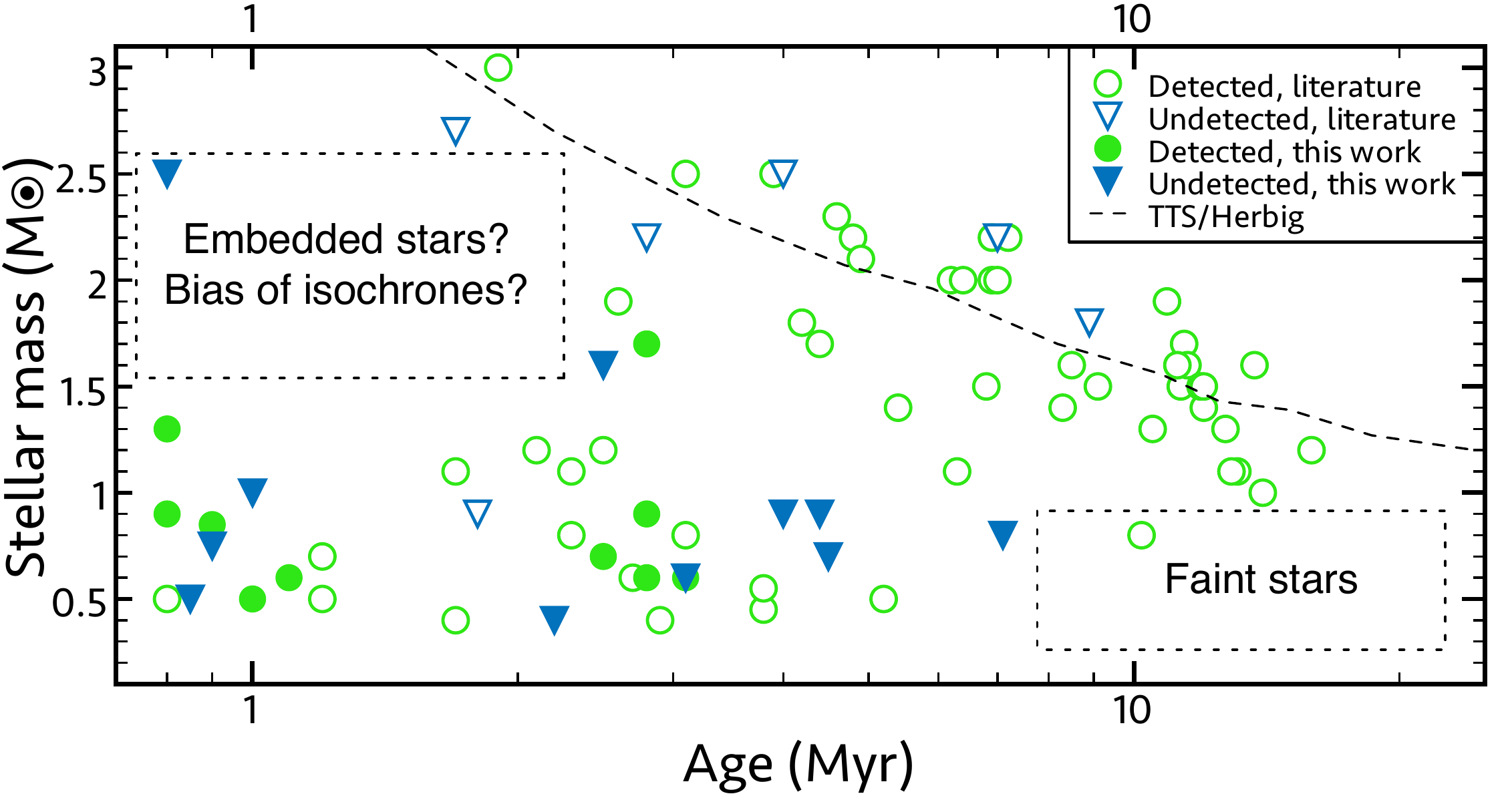} 
  \includegraphics[width=9cm]{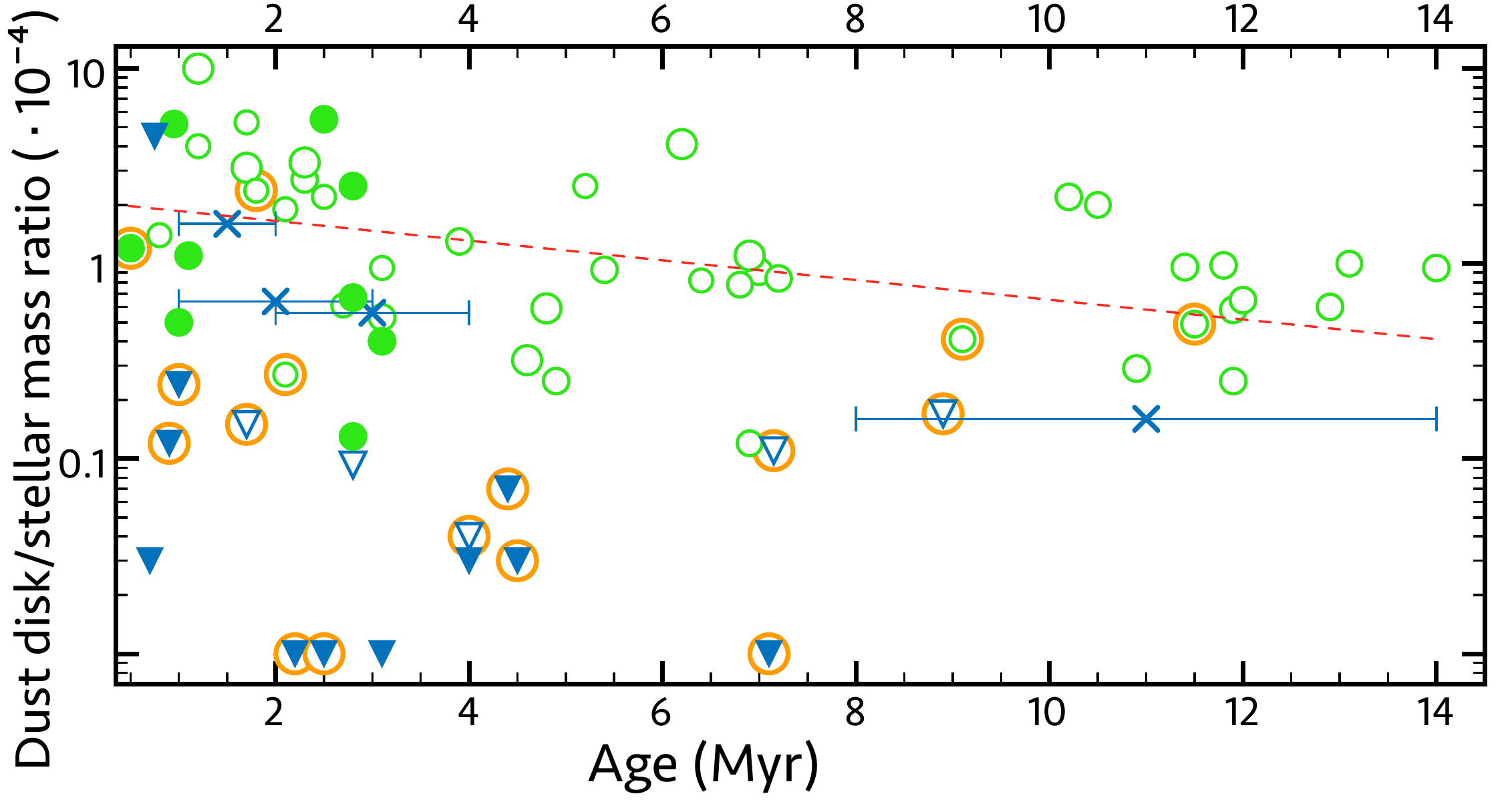} 
     \caption{Stellar and disk mass with time. \textit{Left panel:} Stellar mass for the sample of this work compared with the literature sample analyzed by \citet{Garufi2018}. The dashed line is the formal threshold of G0-type stars. \textit{Right panel:} Same as the left panel for the dust disk mass normalized to the stellar mass. The blue crosses indicate the median ages for the Taurus, Lupus, Chamaeleon, and Upper Sco regions. The orange circles denote the presence of a visual companion at a few dozen au from that target. The dashed line is the fit to the disks that were available before this work.}
          \label{Mass_time}
 \end{figure*}

To comment on the new findings of this work, we first need to place our sample into context. This is done in Fig.\,\ref{Mass_time} by comparing the stellar and disk masses of our sources with those of all other sources with published observations in polarized light. As mentioned in Sect.\,\ref{Stellar_properties}, the targets of this work are in the current framework of PMS tracks on average significantly younger than the literature sample. This is particularly true for nine sources (with age $\sim$1 Myr) that populate a thus far uncovered age interval (see the filled symbols in the left panel of Fig.\,\ref{Mass_time}). The two deserts revealed in this diagram by \citet{Garufi2018} still remain unpopulated. Sub-solar mass stars older than 7$-$8 Myr are not observed in the NIR because of their low luminosity, which prevents them from being observed with current adaptive-optics systems (the only exception is in fact the near TW Hya). The paucity of young super-solar mass stars in the NIR sample is more controversial. \citet{Garufi2018} hypothesized that the prominent envelope that is often associated with these stars (e.g., T Tau or R CrA) could make these targets challenging in the NIR. We also note that some authors \citep[e.g.,][]{Hillenbrand1997} revealed the tendency for higher-mass stars to be older, suggesting an artifact of evolutionary tracks. We nonetheless defer further investigation of this aspect to future works. Another finding that can be read from the diagram in question is that disk detections and non-detections from this sample are equally distributed across stellar mass and time, although it is clear from the older sample from the literature that disks around old stars are always detected.  

The other peculiarity of the targets of this work is their low dust disk mass. Keeping in mind the large uncertainties on these estimates (see Sect.\,\ref{Stellar_properties}), we found that nearly half of the sample is in the low tail of the mass distribution (<\,10 M$_\oplus$) provided by the literature sample (see the right panel of Fig.\,\ref{Mass_time}). The comparison of the literature sample with this dataset also clearly shows that the apparent shallow trend found by \citet{Garufi2018} is an observational bias due to our tendency to observe thus far the most massive disks with a slow mass dispersal (>10 Myr). Unfortunately, this new low-mass population of disks remains largely unresolved (see Sect.\,\ref{Discussion_non-detections}), and their older counterpart is still unobserved.

With this data release, the number of TTSs imaged in the NIR is approximately three times larger than that of Herbig stars. Nonetheless, the characterization of Herbig disks has thus far been better because of their typically higher brightness in scattered light. By definition, TTS disks are on average younger than Herbig disks (see Fig.\,\ref{Mass_time}), and young disks are commonly fainter than old disks \citep{Garufi2018}. The 11 new detections of TTS disks from this work therefore represent a significant boost in the effort of comparing disks at different evolutionary stages.

\subsection{Substructures of TTS disks} \label{Discussion_sub-structures}
The observations we presented support the trend that spiral arms and shadows are primarily {associated with Herbig stars}. We found no convincing evidence of actual shadows when we consider that the azimuthally localized dips seen around V1094 Sco and J1609-1908 are likely due to the scattering properties (being seen along the minor axis) or to an imperfect correction for stellar polarization (see Fig.\,\ref{Stellar_pol}). We did detect a spiral arm (around \mbox{WW Cha}), but it is morphologically associated with the large-scale filaments (see Sect.\,\ref{Individual}), which suggests a different origin from those seen in older Herbig disks \citep[e.g.,][]{Muto2012, Benisty2015}. 

A ring-like structure is revealed in only one disk (around J1609-1908), while it is surprisingly undetected in disks with rings detected by ALMA (V1094 Sco, DoAr 25, and possibly SR4, see Sect.\,\ref{Millimeter_images}). This paucity cannot be fully explained by the general faintness of young disks in scattered light because these disks are bright, nor by the large inclination because disk substructures are seen in more inclined disks \citep[e.g., RY Lup and MY Lup,][]{Langlois2018, Avenhaus2018}. It is tempting to conclude that in young disks the formation of annular gaps in the micron-sized dust grains are less efficient because several older disks are known to host rings in the NIR, similarly to the millimeter \citep[e.g., TW Hya,][]{Andrews2016, vanBoekel2017}. In this regard, it is intriguing that the only ring-like disk in our sample is the oldest detected disk (J1609-1908 with $\sim$3 Myr, to be compared with the median 1.5 Myr of the other detections).

None of our images show any large disk cavity, as recurrently seen in older disks \citep[e.g.,][]{Mayama2012, Avenhaus2014a}. This is in principle at odds with the low NIR excess measured from their SED (as described in Sect.\,\ref{Stellar_properties}), which is lower than 12\% of the stellar flux for 75\% of the sample in a framework where a full disk shows an NIR excess of 15$-$20\% \citep{Banzatti2018}. This most likely indicates that the disk cavities are actually present, but that their size is systematically smaller than the coronagraph ($\sim$15 au). {For the three objects with high-resolution ALMA images (DoAr 25, SR4, and HT Lup), we can rule out cavities larger than $\sim$5 au (see Fig.\,\ref{ALMA_SPHERE})}.

An interesting morphology that could be considered frequent in our sample is the existence of two disk regions with different properties, that is,\,an inner disk region that is bright and compact (called core in Fig.\,\ref{Scaled_imagery}), and an outer and fainter disk region that is separated from the inner region by an abrupt brightness drop (exterior emission). This morphology is seen in at least four objects (V1094 Sco, J1609-1908, HT Lup, and SR4). In Sect.\,\ref{Millimeter_images} we showed that in two cases (HT Lup and SR4), the core emission comes from the only region with millimeter continuum emission, while in another case (V1094 Sco), it matches the location where this emission is optically thick. This supports the view in which dramatic changes in the disk {optical depth} also affect the vertical extent of disks, which is traced by scattered-light observations. This effect may be stronger in young disks because their inner regions may be on average more optically thick. 

Finally, filaments are detected around three of our targets, while this type of structure is certainly rare in more evolved stars. Whether the filaments seen in our images are related to the initial or final stages of the disk evolution may be different from source to source. WW Cha sits within a known network of clumpy filaments \citep{Haikala2005} that resemble a fragmenting molecular cloud. Star formation in this cloud, Cha I, is assumed to have recently come to an end \citep[see e.g.,][]{AlvesdeOliveira2014}. This source is also known to be associated with the highly collimated jet HH 915, large outflows, the bow shock structure HH 931, and a large reflection nebula \citep{Bally2006}. J1615-1921 is similarly embedded in the natal cloud \citep[see Sect.\,\ref{ALMA_SPHERE} and][]{Barenfeld2017}, and our filaments reveal a complex connection with the small disk. On the other hand, the connection between the binary stars DoAr 21 \citep{Loinard2008} and the observed filaments may be unrelated to the star formation. The peculiar structure that we observed in scattered light was also imaged by \citet{Jensen2009} in the mid-IR with the Gemini Thermal-Region Camera Spectrograph (T-ReCs). The authors proposed that DoAr 21 no longer possesses a disk and that the appreciable FIR excess (4.4\% from our calculations) as well as the H$_2$ and PAH observed emission might be due to the strong far-UV (FUV) and X-ray luminosity measured from the binary stars. This energetic flux could soon have photoevaporated the disk and created a small-scale photodissociation region while moving through the cloud. Recently, \citet{Curiel2019} claimed through radio interferometry the existence of two brown dwarfs that would have formed from the rapid fragmentation of the disk. 

\begin{figure}
  \centering
 \includegraphics[width=9.2cm]{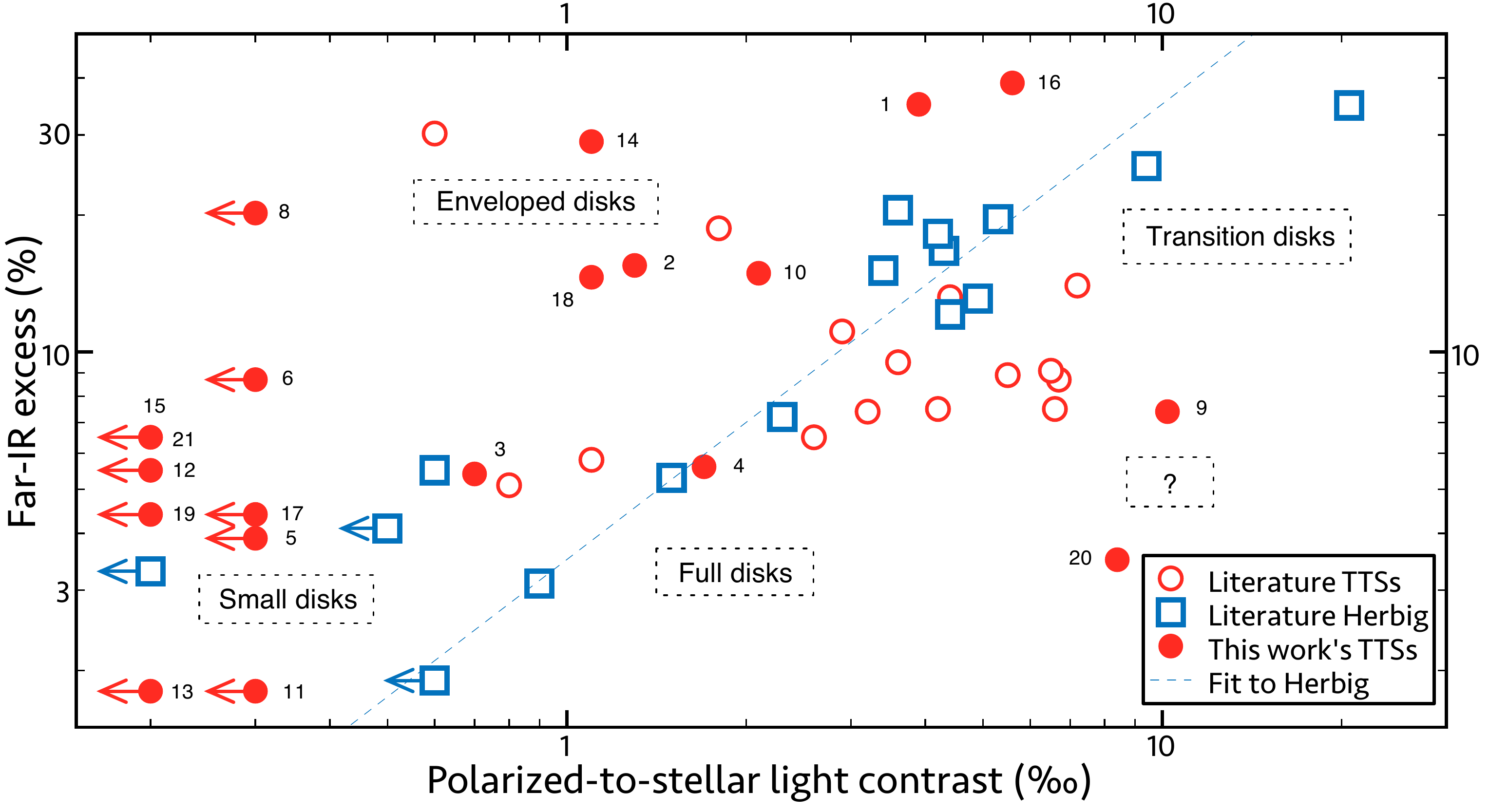} 
     \caption{Far-IR excess vs. disk brightness in scattered light. The dashed line is the fit to the Herbig stars only. Target numbers refer to Table \ref{Sample_properties}.}
          \label{FIR_contrast}
 \end{figure}

\subsection{IR excess of TTS disks}
Large discrepancies between TTS and Herbig disks are also evident when we consider the relation between FIR excess and disk brightness in scattered light. In Herbig disks, these two quantities are always correlated \citep{Garufi2017} and naturally divide transition disks (with bright outer regions) and full disks (with faint outer regions). This correlation arises because the two types of emission originate from the disk surface at several au from the star. From Fig.\,\ref{FIR_contrast}, we see, however, that the trend disappears for TTSs and in particular for the sources of this work. Five detected disks lie above the trend determined for Herbig disks. This behavior can likely be explained by the presence of a remnant envelope around these young sources that only contribute to the FIR budget.  WW Cha (number 1) and J1615-1921 (number 16), which are imaged with extended filaments in Fig.\,\ref{Imagery}, sit in this region of the diagram. Most of our non-detections lie in a parameter range that is typical of small disks (see Sect.\,\ref{Discussion_non-detections}). In these disks, the two quantities may still correlate, but the polarized light is, unlike the FIR excess, not accessible because of the coronagraph.

Other outliers such as HK Lup (number 8, see Sect.\,\ref{Discussion_non-detections}), V1094 Sco (number 9), and DoAr 25 (number 20) are less intuitively explained. The case of DoAr 25 is particularly intriguing because other disks with similarly low FIR excess (3.5\%) are not even detected, whereas our NIR map shows one of the most prominent disks ever imaged. \citet{Andrews2008} noted the very low IR excess of this source given the bright millimeter emission and proposed that the material therein is in an advanced state of evolution. Our image does not provide a qualitative confirmation to this idea given the efficient scattering of $\text{micron}$-sized dust grains and the obviously flared structure of the disk. We currently do not have any explanation for this conundrum, but note that a morphologically similar disk \citep[IM Lup,][]{Avenhaus2018} has an analogously low FIR excess (7.5\%).

\subsection{Undetected disks} \label{Discussion_non-detections}
The main images of Fig.\,\ref{Imagery} show that the fraction of non-detections in our sample is $\sim$50\%. This is clearly larger than that from published observations ($\sim$10\%) because the general tendency is to avoid publication of non-detections and because our strategy was to also observe low-mass disks. Figure \ref{Mass_time} clearly shows that all but one of our non-detections have a low dust disk mass (<\,10 M$_\oplus$), and their small size ($\lesssim20$ au in radius) probably explains why they were not detected in the $Q_\phi$ image. This claim is also supported by the comparison with the known millimeter sizes, as shown in Fig.\,\ref{Outer_radii}.

As of today, the lowest mass disks ever resolved in polarized light are DZ Cha \citep{Canovas2018} and AK Sco \citep{Garufi2017}, which  have a dust mass of <\,3 M$_\oplus$ and 4 M$_\oplus$, respectively. From this work, J1615-1921 also has a very low-mass disk (4 M$_\oplus$), but our image mostly recovers light from the surrounding medium. In addition to resolved maps, polarized signal from a small disk can be recovered from unresolved regions, as described in Sect.\,\ref{Stellar_polarization}. In this regard, the unresolved detections of J1603-2031 and VV Sco (with masses of 1 M$_\oplus$ and 2 M$_\oplus$, respectively) are the first of their kind as no disk with a lower mass has ever been detected in scattered light. Clearly, the employment of this technique on available datasets from the literature may yield further unresolved detections.

As mentioned in Sect.\,\ref{Individual}, as many as seven undetected low-mass disks have a visual stellar companion within a few hundred au. Here we assume that these disks have an outer radius in small dust grains $r_{\rm d}$ of about 15 au, as suggested by the unresolved nature of the detected scattered light (see Sect.\,\ref{Stellar_polarization}). This is expected based on the average disk sizes measured from ALMA surveys \citep[less than $\sim$20 au in radius, see, e.g.,][]{Barenfeld2017, Cieza2019}. From our sample, the ratios between $r_{\rm d}$  and the apparent separation of the companion $a_{\rm c}$ would span from 0.1 to 0.5 with a median value of $r_{\rm d}/a_{\rm c}=0.2$. Furthermore, these values are only upper limits because the distance of the companion that we measure is only the projected separation. Therefore, most if not all targets are inconsistent with the $r_{\rm d}/a_{\rm c}$ ratio that was analytically calculated in case of tidal truncation by a companion on a circular orbit \citep[0.3,][]{Artymowicz1994}. This prediction applies to the gaseous disk, but our observations are expected to indirectly trace the same component (see HT Lup in Fig.\,\ref{ALMA_SPHERE}) because of the dynamical coupling between small dust grains and gas. 

The inconsistency between the observed and predicted $r_{\rm d}/a_{\rm c}$ ratio has previously been pointed out by some authors \citep[see, e.g.,][]{Manara2019}, although their $r_{\rm d}$ is typically constrained through millimeter continuum emission, which allows invoking an anomalously high gas-to-dust size ratio to reconcile observations and simulations. Instead, our upper limit on $r_{\rm d}$ is related to the gaseous extent. It is still possible that some disks are moderately larger than 15 au and that the outer regions are not directly illuminated (see below the case of HK Lup). However, it is unlikely that this explanation holds for all disks given the high incidence of small cavities suggested by the SED (see Sect.\,\ref{Discussion_sub-structures}). For statistical reasons, it is also very unlikely that each of the companions has an eccentricity that is high enough to allow a much closer passage at periastron \citep[see][]{Duchene2013, Manara2019}. All these arguments may therefore suggest that the tidal truncation is not (always) responsible for the systematic non-detections of disks in large binary systems and that a major role in limiting the disk size is played by the formation history of disks and not their  evolution.   

Everything said in this section, however, does not apply to one target, HK Lup (number 8). Figures \ref{Outer_radii}, \ref{Mass_time}, and \ref{FIR_contrast} clearly show that this non-detection is inconsistent with a small disk given its size in the millimeter, dust mass, and FIR excess. Thus, this non-detection is more likely explained by a self-shadowing disk, that is, a disk where the puffed-up inner regions can cast a shadow cone outward and prevent the outer disk from being illuminated \citep{Dullemond2001}. Disks partly showing this effect are known \citep[e.g., HD163296,][]{Garufi2014b, MuroArena2018}, but a shadowing as efficient as this has not been reported before. As commented in Sect.\,\ref{Stellar_polarization}, the inner disk of HK Lup is indeed detected. This indicates that a significant fraction of the stellar light is scattered off by this inner disk component (of a few au in size), and that this may be the primary responsible factor for the shadow cast outward.

\section{Summary and conclusions} \label{Conclusions}
This paper is the second release of data from the DARTTS program. The sample consists of 21 new NIR polarimetric images of protoplanetary disks, making it the largest sample ever released for this type of data. Scattered light is confidently detected around 14 of the 21 targeted TTSs. For 11 of these, the polarized flux is immediately recovered from their map, while for the remaining 3 we were only able to detect some unresolved signal, with polarization vectors oriented coherently with the known disk geometry. Other unresolved disks may be detected but require future confirmation from millimeter images. The main results of our demographical analysis are listed below.
\begin{itemize}
\item The 11 imaged disks have brightnesses spanning a factor 15 and apparent sizes spanning a factor 7 (from 50 au to 320 au). Sizes are on average 50\% larger than those constrained by ALMA with continuum millimeter emission.

\item Bright visual companions are visible around eight targets. With one notable exception (HT Lup), the presence of a companion implies a disk non-detection. These are best explained by the disk being smaller than 15 au. In most cases, this is significantly less than one-third of the separation of the companion, implying that the tidal truncation may not (always) be the cause of their small size.

\item The sample is young on average. The median age of all targets constrained from stellar tracks is $\sim$2 Myr, which is significantly younger than the age of all disks with published observations in scattered light \mbox{($\sim$7 Myr)}.  

\item The sample alleviates the bias for which only massive planet-forming disks are imaged in scattered light. Half of our disks are the least massive disks ever observed. The dust mass of the disk around J1603-2031 ($\sim$1 M$_\oplus$) represents a new benchmark. 

\item The paucity of shadows and spirals in the sample supports the trend for which these features are associated with older disks {and/or with more massive stars}.

\item Prominent ring-like structures detected with ALMA also remain elusive in some of our images (particularly in V1094 Sco and DoAr 25). This contrasts with what is seen in older disks and may intuitively mean that initially (<2$-$3 Myr), disk gaps are preferentially sculpted for large grains only.  

\item Although the SEDs of many targets show a lower NIR excess, none of our images reveals a disk cavity. These could indicate a population of cavities smaller than 15 au or that small dust grains still linger in the cavities.

\item Four disks show a bright core emission that is discontinued from an outer tenuous region by an abrupt brightness drop. ALMA images show that the large dust is either confined within this bright core (HT Lup and SR4) or that the millimeter emission within this region is optically thick (V1094 Sco).

\item Three objects (WW Cha, J1615-1921, and DoAr 21) show extended filaments, which is  indicative of peculiar processes in the immediate surrounding, such as disk accretion or photoevaporation.

\end{itemize}

In addition to the demographical approach, some of the 29 DARTTS sources presented here and in \citet{Avenhaus2018} certainly deserve a dedicated analysis. Of the data release of this work, we mention DoAr 25 (showing a spectacular disk despite its modest IR excess), WW Cha (providing an exceptional laboratory for the study of the disk interaction with the medium), HT Lup (hosting the only disk with a stellar companion that is visually embedded), and HK Lup (having the most massive undetected disk because of very efficient self-shadowing). This work also increased the number of disks for which both NIR polarimetric and ALMA high-resolution (<0.1\arcsec) images are available. With V1094 Sco, DoAr 25, SR4, and HT Lup, this number now amounts to a couple of dozens. This comparison is of fundamental importance to determine how dust differentiation induced by ongoing planet formation changes throughout disk evolution, and this will also be addressed by DARTTS-A (P.I.: S.\,Perez). 

The immediate goal of the community studying NIR disk imaging should be to provide a more complete and less biased census of planet-forming disks, following the strategy of the ALMA community. The DARTTS-S program represents a first important step in this direction, having provided access to younger, fainter, and smaller disks. Furthermore, the improvement of data reduction pipelines \citep[e.g., IRDAP,][]{vanHolstein2019} and of our view of less exceptional disks will help us define the best observing strategy in future campaigns, such as the SPHERE large program DESTINYS (P.I.: C.\,Ginski). 

\begin{acknowledgements}
     {We thank the referee for comments which helped to improve the paper.} We are grateful to S.E.\ van Terwisga and to S.\ Andrews and the DSHARP team for making available their ALMA images, as well as to E.\ Pancino for useful discussions. We also thank the ESO technical operators at the Paranal observatory for their valuable help during the observations. This work has been supported by the project PRIN-INAF 2016 The Cradle of Life - GENESIS-SKA (General Conditions in Early Planetary Systems for the rise of life with SKA). We also acknowledge support from INAF/Frontiera (Fostering high ResolutiON Technology and Innovation for Exoplanets and Research in Astrophysics) through the "Progetti Premiali" funding scheme of the Italian Ministry of Education, University, and Research. SP acknowledges support from CONICYT-FONDECYT grant \#1191934. LAC was supported by CONICYT-FONDECYT grant number 1171246. GHMB acknowledges support from the European Research Council (ERC) under the European Union's Horizon 2020 research and innovation programme (grand agreement no.\ 757957). GvdP acknowledges funding from ANR of France (ANR-16-CE310013). AZu acknowledges support from the CONICYT + PAI/ Convocatoria nacional subvenci\'on a la instalaci\'on en la academia, convocatoria 2017 + Folio PAI77170087. SPHERE is an instrument designed and built by a consortium consisting of IPAG (Grenoble, France), MPIA (Heidelberg, Germany), LAM (Marseille, France), LESIA (Paris, France), Laboratoire Lagrange (Nice, France), INAF - Osservatorio di Padova (Italy), Observatoire de Geneve (Switzerland), ETH Zurich (Switzerland), NOVA (Netherlands), ONERA (France), and ASTRON (Netherlands) in collaboration with ESO. SPHERE was funded by ESO, with additional contributions from the CNRS (France), MPIA (Germany), INAF (Italy), FINES (Switzerland), and NOVA (Netherlands). SPHERE also received funding from the European Commission Sixth and Seventh Framework Programs as part of the Optical Infrared Coordination Network for Astronomy (OPTICON) under grant number RII3-Ct-2004-001566 for FP6 (2004-2008), grant number 226604 for FP7 (2009-2012), and grant number 312430 for FP7 (2013-2016). This work has made use of the SIMBAD database, operated at the CDS, Strasbourg, France, as well as of data from the European Space Agency (ESA) mission
{\it Gaia} (\url{https://www.cosmos.esa.int/gaia}), processed by the {\it Gaia} Data Processing and Analysis Consortium (DPAC,
\url{https://www.cosmos.esa.int/web/gaia/dpac/consortium}). Funding for the DPAC has been provided by national institutions, in particular the institutions participating in the {\it Gaia} Multilateral Agreement.
     \end{acknowledgements}

\bibliographystyle{aa} 
\bibliography{../MasterReference.bib} 

\begin{thebibliography}{81}
\expandafter\ifx\csname natexlab\endcsname\relax\def\natexlab#1{#1}\fi

\bibitem[{{Alves de Oliveira} {et~al.}(2014){Alves de Oliveira}, {Schneider},
  {Mer{\'\i}n}, {Prusti}, {Ribas}, {Cox}, {Vavrek}, {K{\"o}nyves},
  {Arzoumanian}, {Puga}, {Pilbratt}, {K{\'o}sp{\'a}l}, {Andr{\'e}}, {Didelon},
  {Men'shchikov}, {Royer}, {Waelkens}, {Bontemps}, {Winston}, \&
  {Spezzi}}]{AlvesdeOliveira2014}
{Alves de Oliveira}, C., {Schneider}, N., {Mer{\'\i}n}, B., {et~al.} 2014,
  \aap, 568, A98

\bibitem[{{Andrews} {et~al.}(2018){Andrews}, {Huang}, {P{\'e}rez}, {Isella},
  {Dullemond}, {Kurtovic}, {Guzm{\'a}n}, {Carpenter}, {Wilner}, \&
  {Zhang}}]{Andrews2018}
{Andrews}, S.~M., {Huang}, J., {P{\'e}rez}, L.~M., {et~al.} 2018, \apj, 869,
  L41

\bibitem[{{Andrews} {et~al.}(2008){Andrews}, {Hughes}, {Wilner}, \&
  {Qi}}]{Andrews2008}
{Andrews}, S.~M., {Hughes}, A.~M., {Wilner}, D.~J., \& {Qi}, C. 2008, \apjl,
  678, L133

\bibitem[{{Andrews} \& {Williams}(2007)}]{Andrews2007}
{Andrews}, S.~M. \& {Williams}, J.~P. 2007, \apj, 671, 1800

\bibitem[{{Andrews} {et~al.}(2016){Andrews}, {Wilner}, {Zhu}, {Birnstiel},
  {Carpenter}, {P{\'e}rez}, {Bai}, {{\"O}berg}, {Hughes}, {Isella}, \&
  {Ricci}}]{Andrews2016}
{Andrews}, S.~M., {Wilner}, D.~J., {Zhu}, Z., {et~al.} 2016, \apjl, 820, L40

\bibitem[{{Ansdell} {et~al.}(2018){Ansdell}, {Williams}, {Trapman}, {van
  Terwisga}, {Facchini}, {Manara}, {van der Marel}, {Miotello}, {Tazzari},
  {Hogerheijde}, {Guidi}, {Testi}, \& {van Dishoeck}}]{Ansdell2018}
{Ansdell}, M., {Williams}, J.~P., {Trapman}, L., {et~al.} 2018, \apj, 859, 21

\bibitem[{{Ansdell} {et~al.}(2016){Ansdell}, {Williams}, {van der Marel},
  {Carpenter}, {Guidi}, {Hogerheijde}, {Mathews}, {Manara}, {Miotello},
  {Natta}, {Oliveira}, {Tazzari}, {Testi}, {van Dishoeck}, \& {van
  Terwisga}}]{Ansdell2016}
{Ansdell}, M., {Williams}, J.~P., {van der Marel}, N., {et~al.} 2016, \apj,
  828, 46

\bibitem[{{Anthonioz} {et~al.}(2015){Anthonioz}, {M{\'e}nard}, {Pinte}, {Le
  Bouquin}, {Benisty}, {Thi}, {Absil}, {Duch{\^e}ne}, {Augereau}, {Berger},
  {Casassus}, {Duvert}, {Lazareff}, {Malbet}, {Millan-Gabet}, {Schreiber},
  {Traub}, \& {Zins}}]{Anthonioz2015}
{Anthonioz}, F., {M{\'e}nard}, F., {Pinte}, C., {et~al.} 2015, \aap, 574, A41

\bibitem[{{Artymowicz} \& {Lubow}(1994)}]{Artymowicz1994}
{Artymowicz}, P. \& {Lubow}, S.~H. 1994, \apj, 421, 651

\bibitem[{{Avenhaus} {et~al.}(2018){Avenhaus}, {Quanz}, {Garufi}, {Perez},
  {Casassus}, {Pinte}, {Bertrang}, {Caceres}, {Benisty}, \&
  {Dominik}}]{Avenhaus2018}
{Avenhaus}, H., {Quanz}, S.~P., {Garufi}, A., {et~al.} 2018, The Astrophysical
  Journal, 863, 44

\bibitem[{{Avenhaus} {et~al.}(2014){Avenhaus}, {Quanz}, {Schmid}, {Meyer},
  {Garufi}, {Wolf}, \& {Dominik}}]{Avenhaus2014a}
{Avenhaus}, H., {Quanz}, S.~P., {Schmid}, H.~M., {et~al.} 2014, \apj, 781, 87

\bibitem[{{Bally} {et~al.}(2006){Bally}, {Walawender}, {Luhman}, \&
  {Fazio}}]{Bally2006}
{Bally}, J., {Walawender}, J., {Luhman}, K.~L., \& {Fazio}, G. 2006, \aj, 132,
  1923

\bibitem[{{Banzatti} {et~al.}(2018){Banzatti}, {Garufi}, {Kama}, {Benisty},
  {Brittain}, {Pontoppidan}, \& {Rayner}}]{Banzatti2018}
{Banzatti}, A., {Garufi}, A., {Kama}, M., {et~al.} 2018, \aap, 609, L2

\bibitem[{{Baraffe} {et~al.}(2015){Baraffe}, {Homeier}, {Allard}, \&
  {Chabrier}}]{Baraffe2015}
{Baraffe}, I., {Homeier}, D., {Allard}, F., \& {Chabrier}, G. 2015, \aap, 577,
  A42

\bibitem[{{Barenfeld} {et~al.}(2016){Barenfeld}, {Carpenter}, {Ricci}, \&
  {Isella}}]{Barenfeld2016}
{Barenfeld}, S.~A., {Carpenter}, J.~M., {Ricci}, L., \& {Isella}, A. 2016,
  \apj, 827, 142

\bibitem[{{Barenfeld} {et~al.}(2017){Barenfeld}, {Carpenter}, {Sargent},
  {Isella}, \& {Ricci}}]{Barenfeld2017}
{Barenfeld}, S.~A., {Carpenter}, J.~M., {Sargent}, A.~I., {Isella}, A., \&
  {Ricci}, L. 2017, \apj, 851, 85

\bibitem[{{Benisty} {et~al.}(2015){Benisty}, {Juhasz}, {Boccaletti},
  {Avenhaus}, {Milli}, {Thalmann}, {Dominik}, {Pinilla}, {Buenzli}, {Pohl},
  {Beuzit}, {Birnstiel}, {de Boer}, {Bonnefoy}, {Chauvin}, {Christiaens},
  {Garufi}, {Grady}, {Henning}, {Huelamo}, {Isella}, {Langlois}, {M{\'e}nard},
  {Mouillet}, {Olofsson}, {Pantin}, {Pinte}, \& {Pueyo}}]{Benisty2015}
{Benisty}, M., {Juhasz}, A., {Boccaletti}, A., {et~al.} 2015, \aap, 578, L6

\bibitem[{{Bertrang} {et~al.}(2018){Bertrang}, {Avenhaus}, {Casassus},
  {Montesinos}, {Kirchschlager}, {Perez}, {Cieza}, \& {Wolf}}]{Bertrang2018}
{Bertrang}, G.~H.-M., {Avenhaus}, H., {Casassus}, S., {et~al.} 2018, \mnras,
  474, 5105

\bibitem[{{Beuzit} {et~al.}(2008){Beuzit}, {Feldt}, {Dohlen}, {Mouillet},
  {Puget}, {Wildi}, {Abe}, {Antichi}, {Baruffolo}, {Baudoz}, {Boccaletti},
  {Carbillet}, {Charton}, {Claudi}, {Downing}, {Fabron}, {Feautrier},
  {Fedrigo}, {Fusco}, {Gach}, {Gratton}, {Henning}, {Hubin}, {Joos}, {Kasper},
  {Langlois}, {Lenzen}, {Moutou}, {Pavlov}, {Petit}, {Pragt}, {Rabou}, {Rigal},
  {Roelfsema}, {Rousset}, {Saisse}, {Schmid}, {Stadler}, {Thalmann}, {Turatto},
  {Udry}, {Vakili}, \& {Waters}}]{Beuzit2008}
{Beuzit}, J.-L., {Feldt}, M., {Dohlen}, K., {et~al.} 2008, in Society of
  Photo-Optical Instrumentation Engineers (SPIE) Conference Series, Vol. 7014,
  Society of Photo-Optical Instrumentation Engineers (SPIE) Conference Series,
  18

\bibitem[{{Beuzit} {et~al.}(2019){Beuzit}, {Vigan}, {Mouillet}, {Dohlen},
  {Gratton}, {Boccaletti}, {Sauvage}, {Schmid}, {Langlois}, {Petit},
  {Baruffolo}, {Feldt}, {Milli}, {Wahhaj}, {Abe}, {Anselmi}, {Antichi},
  {Barette}, {Baudrand}, {Baudoz}, {Bazzon}, {Bernardi}, {Blanchard}, {Brast},
  {Bruno}, {Buey}, {Carbillet}, {Carle}, {Cascone}, {Chapron}, {Chauvin},
  {Charton}, {Claudi}, {Costille}, {De Caprio}, {Delboulb{\'e}}, {Desidera},
  {Dominik}, {Downing}, {Dupuis}, {Fabron}, {Fantinel}, {Farisato},
  {Feautrier}, {Fedrigo}, {Fusco}, {Gigan}, {Ginski}, {Girard}, {Giro},
  {Gisler}, {Gluck}, {Gry}, {Henning}, {Hubin}, {Hugot}, {Incorvaia}, {Jaquet},
  {Kasper}, {Lagadec}, {Lagrange}, {Coroller}, {Mignant}, {Ruyet}, {Lessio},
  {Lizon}, {Llored}, {Lundin}, {Madec}, {Magnard}, {Marteaud}, {Martinez},
  {Maurel}, {M{\'e}nard}, {Mesa}, {M{\"o}ller-Nilsson}, {Moulin}, {Moutou},
  {Orign{\'e}}, {Parisot}, {Pavlov}, {Perret}, {Pragt}, {Puget}, {Rabou},
  {Ramos}, {Reess}, {Rigal}, {Rochat}, {Roelfsema}, {Rousset}, {Roux},
  {Saisse}, {Salasnich}, {Santambrogio}, {Scuderi}, {Segransan}, {Sevin},
  {Siebenmorgen}, {Soenke}, {Stadler}, {Suarez}, {Tiph{\`e}ne}, {Turatto},
  {Udry}, {Vakili}, {Waters}, {Weber}, {Wildi}, {Zins}, \&
  {Zurlo}}]{Beuzit2019}
{Beuzit}, J.~L., {Vigan}, A., {Mouillet}, D., {et~al.} 2019, arXiv e-prints,
  arXiv:1902.04080

\bibitem[{{Birnstiel} {et~al.}(2018){Birnstiel}, {Dullemond}, {Zhu}, {Andrews},
  {Bai}, {Wilner}, {Carpenter}, {Huang}, {Isella}, {Benisty}, {P{\'e}rez}, \&
  {Zhang}}]{Birnstiel2018}
{Birnstiel}, T., {Dullemond}, C.~P., {Zhu}, Z., {et~al.} 2018, \apjl, 869, L45

\bibitem[{{Bouvier} \& {Appenzeller}(1992)}]{Bouvier1992}
{Bouvier}, J. \& {Appenzeller}, I. 1992, \aaps, 92, 481

\bibitem[{{Bouy} \& {Mart{\'\i}n}(2009)}]{Bouy2009}
{Bouy}, H. \& {Mart{\'\i}n}, E.~L. 2009, \aap, 504, 981

\bibitem[{{Bressan} {et~al.}(2012){Bressan}, {Marigo}, {Girardi}, {Salasnich},
  {Dal Cero}, {Rubele}, \& {Nanni}}]{Bressan2012}
{Bressan}, A., {Marigo}, P., {Girardi}, L., {et~al.} 2012, \mnras, 427, 127

\bibitem[{{Canovas} {et~al.}(2015){Canovas}, {M{\'e}nard}, {de Boer}, {Pinte},
  {Avenhaus}, \& {Schreiber}}]{Canovas2015}
{Canovas}, H., {M{\'e}nard}, F., {de Boer}, J., {et~al.} 2015, \aap, 582, L7

\bibitem[{{Canovas} {et~al.}(2018){Canovas}, {Montesinos}, {Schreiber},
  {Cieza}, {Eiroa}, {Meeus}, {de Boer}, {M{\'e}nard}, {Wahhaj},
  {Riviere-Marichalar}, {Olofsson}, {Garufi}, {Rebollido}, {van Holstein},
  {Caceres}, {Hardy}, \& {Villaver}}]{Canovas2018}
{Canovas}, H., {Montesinos}, B., {Schreiber}, M.~R., {et~al.} 2018, \aap, 610,
  A13

\bibitem[{{Carbillet} {et~al.}(2011){Carbillet}, {Bendjoya}, {Abe}, {Guerri},
  {Boccaletti}, {Daban}, {Dohlen}, {Ferrari}, {Robbe-Dubois}, {Douet}, \&
  {Vakili}}]{Carbillet2011}
{Carbillet}, M., {Bendjoya}, P., {Abe}, L., {et~al.} 2011, Experimental
  Astronomy, 30, 39

\bibitem[{{Casassus} {et~al.}(2013){Casassus}, {van der Plas}, {M}, {Dent},
  {Fomalont}, {Hagelberg}, {Hales}, {Jord{\'a}n}, {Mawet}, {M{\'e}nard},
  {Wootten}, {Wilner}, {Hughes}, {Schreiber}, {Girard}, {Ercolano}, {Canovas},
  {Rom{\'a}n}, \& {Salinas}}]{Casassus2013}
{Casassus}, S., {van der Plas}, G., {M}, S.~P., {et~al.} 2013, \nat, 493, 191

\bibitem[{{Choi} {et~al.}(2016){Choi}, {Dotter}, {Conroy}, {Cantiello},
  {Paxton}, \& {Johnson}}]{Choi2016}
{Choi}, J., {Dotter}, A., {Conroy}, C., {et~al.} 2016, \apj, 823, 102

\bibitem[{{Cieza} {et~al.}(2007){Cieza}, {Padgett}, {Stapelfeldt}, {Augereau},
  {Harvey}, {Evans}, {Mer{\'\i}n}, {Koerner}, {Sargent}, {van Dishoeck},
  {Allen}, {Blake}, {Brooke}, {Chapman}, {Huard}, {Lai}, {Mundy}, {Myers},
  {Spiesman}, \& {Wahhaj}}]{Cieza2007}
{Cieza}, L., {Padgett}, D.~L., {Stapelfeldt}, K.~R., {et~al.} 2007, \apj, 667,
  308

\bibitem[{{Cieza} {et~al.}(2019){Cieza}, {Ru{\'\i}z-Rodr{\'\i}guez}, {Hales},
  {Casassus}, {P{\'e}rez}, {Gonzalez-Ruilova}, {C{\'a}novas}, {Williams},
  {Zurlo}, {Ansdell}, {Avenhaus}, {Bayo}, {Bertrang}, {Christiaens}, {Dent},
  {Ferrero}, {Gamen}, {Olofsson}, {Orcajo}, {Pe{\~n}a Ram{\'\i}rez},
  {Principe}, {Schreiber}, \& {van der Plas}}]{Cieza2019}
{Cieza}, L.~A., {Ru{\'\i}z-Rodr{\'\i}guez}, D., {Hales}, A., {et~al.} 2019,
  \mnras, 482, 698

\bibitem[{{Curiel} {et~al.}(2019){Curiel}, {Ortiz-Le{\'o}n}, {Mioduszewski}, \&
  {Torres}}]{Curiel2019}
{Curiel}, S., {Ortiz-Le{\'o}n}, G.~N., {Mioduszewski}, A.~J., \& {Torres},
  R.~M. 2019, \apj, 884, 13

\bibitem[{{de Boer} {et~al.}(2019){de Boer}, {Langlois}, {van Holstein},
  {Girard}, {Mouillet}, {Vigan}, {Dohlen}, {Snik}, {Keller}, {Ginski}, {Stam},
  {Milli}, {Wahhaj}, {Kasper}, {Schmid}, {Rabou}, {Gluck}, {Hugot}, {Perret},
  {Martinez}, {Weber}, {Pragt}, {Sauvage}, {Boccaletti}, {Le Coroller},
  {Dominik}, {Henning}, {Lagadec}, {M{\'e}nard}, {Turatto}, {Udry}, {Chauvin},
  {Feldt}, \& {Beuzit}}]{deBoer2019}
{de Boer}, J., {Langlois}, M., {van Holstein}, R.~G., {et~al.} 2019, arXiv
  e-prints, arXiv:1909.13107

\bibitem[{{Dohlen} {et~al.}(2008){Dohlen}, {Langlois}, {Saisse}, {Hill},
  {Origne}, {Jacquet}, {Fabron}, {Blanc}, {Llored}, {Carle}, {Moutou}, {Vigan},
  {Boccaletti}, {Carbillet}, {Mouillet}, \& {Beuzit}}]{Dohlen2008}
{Dohlen}, K., {Langlois}, M., {Saisse}, M., {et~al.} 2008, in Society of
  Photo-Optical Instrumentation Engineers (SPIE) Conference Series, Vol. 7014,
  Society of Photo-Optical Instrumentation Engineers (SPIE) Conference Series,
  3

\bibitem[{{Duch{\^e}ne} \& {Kraus}(2013)}]{Duchene2013}
{Duch{\^e}ne}, G. \& {Kraus}, A. 2013, \araa, 51, 269

\bibitem[{{Dullemond} {et~al.}(2001){Dullemond}, {Dominik}, \&
  {Natta}}]{Dullemond2001}
{Dullemond}, C.~P., {Dominik}, C., \& {Natta}, A. 2001, \apj, 560, 957

\bibitem[{{Fedele} {et~al.}(2017){Fedele}, {Carney}, {Hogerheijde}, {Walsh},
  {Miotello}, {Klaassen}, {Bruderer}, {Henning}, \& {van
  Dishoeck}}]{Fedele2017}
{Fedele}, D., {Carney}, M., {Hogerheijde}, M.~R., {et~al.} 2017, \aap, 600, A72

\bibitem[{{Gaia Collaboration} {et~al.}(2018){Gaia Collaboration}, {Brown},
  {Vallenari}, {Prusti}, {de Bruijne}, {Babusiaux}, {Bailer-Jones}, {Biermann},
  {Evans}, {Eyer}, {Jansen}, {Jordi}, {Klioner}, {Lammers}, {Lindegren},
  {Luri}, {Mignard}, {Panem}, {Pourbaix}, {Randich}, {Sartoretti}, {Siddiqui},
  {Soubiran}, {van Leeuwen}, {Walton}, {Arenou}, {Bastian}, {Cropper},
  {Drimmel}, {Katz}, {Lattanzi}, {Bakker}, {Cacciari}, {Casta{\~n}eda},
  {Chaoul}, {Cheek}, {De Angeli}, {Fabricius}, {Guerra}, {Holl}, {Masana},
  {Messineo}, {Mowlavi}, {Nienartowicz}, {Panuzzo}, {Portell}, {Riello},
  {Seabroke}, {Tanga}, {Th{\'e}venin}, {Gracia-Abril}, {Comoretto},
  {Garcia-Reinaldos}, {Teyssier}, {Altmann}, {Andrae}, {Audard},
  {Bellas-Velidis}, {Benson}, {Berthier}, {Blomme}, {Burgess}, {Busso},
  {Carry}, {Cellino}, {Clementini}, {Clotet}, {Creevey}, {Davidson}, {De
  Ridder}, {Delchambre}, {Dell'Oro}, {Ducourant},
  {Fern{\'a}ndez-Hern{\'a}ndez}, {Fouesneau}, {Fr{\'e}mat}, {Galluccio},
  {Garc{\'\i}a-Torres}, {Gonz{\'a}lez-N{\'u}{\~n}ez}, {Gonz{\'a}lez-Vidal},
  {Gosset}, {Guy}, {Halbwachs}, {Hambly}, {Harrison}, {Hern{\'a}ndez},
  {Hestroffer}, {Hodgkin}, {Hutton}, {Jasniewicz}, {Jean-Antoine-Piccolo},
  {Jordan}, {Korn}, {Krone-Martins}, {Lanzafame}, {Lebzelter}, {L{\"o}ffler},
  {Manteiga}, {Marrese}, {Mart{\'\i}n-Fleitas}, {Moitinho}, {Mora}, {Muinonen},
  {Osinde}, {Pancino}, {Pauwels}, {Petit}, {Recio-Blanco}, {Richards},
  {Rimoldini}, {Robin}, {Sarro}, {Siopis}, {Smith}, {Sozzetti}, {S{\"u}veges},
  {Torra}, {van Reeven}, {Abbas}, {Abreu Aramburu}, {Accart}, {Aerts},
  {Altavilla}, {{\'A}lvarez}, {Alvarez}, {Alves}, {Anderson}, {Andrei},
  {Anglada Varela}, {Antiche}, {Antoja}, {Arcay}, {Astraatmadja}, {Bach},
  {Baker}, {Balaguer-N{\'u}{\~n}ez}, {Balm}, {Barache}, {Barata}, {Barbato},
  {Barblan}, {Barklem}, {Barrado}, {Barros}, {Barstow}, {Bartholom{\'e}
  Mu{\~n}oz}, {Bassilana}, {Becciani}, {Bellazzini}, {Berihuete}, {Bertone},
  {Bianchi}, {Bienaym{\'e}}, {Blanco-Cuaresma}, {Boch}, {Boeche}, {Bombrun},
  {Borrachero}, {Bossini}, {Bouquillon}, {Bourda}, {Bragaglia}, {Bramante},
  {Breddels}, {Bressan}, {Brouillet}, {Br{\"u}semeister}, {Brugaletta},
  {Bucciarelli}, {Burlacu}, {Busonero}, {Butkevich}, {Buzzi}, {Caffau},
  {Cancelliere}, {Cannizzaro}, {Cantat-Gaudin}, {Carballo}, {Carlucci},
  {Carrasco}, {Casamiquela}, {Castellani}, {Castro-Ginard}, {Charlot},
  {Chemin}, {Chiavassa}, {Cocozza}, {Costigan}, {Cowell}, {Crifo}, {Crosta},
  {Crowley}, {Cuypers}, {Dafonte}, {Damerdji}, {Dapergolas}, {David}, {David},
  {de Laverny}, {De Luise}, {De March}, {de Martino}, {de Souza}, {de Torres},
  {Debosscher}, {del Pozo}, {Delbo}, {Delgado}, {Delgado}, {Di Matteo},
  {Diakite}, {Diener}, {Distefano}, {Dolding}, {Drazinos}, {Dur{\'a}n},
  {Edvardsson}, {Enke}, {Eriksson}, {Esquej}, {Eynard Bontemps}, {Fabre},
  {Fabrizio}, {Faigler}, {Falc{\~a}o}, {Farr{\`a}s Casas}, {Federici},
  {Fedorets}, {Fernique}, {Figueras}, {Filippi}, {Findeisen}, {Fonti},
  {Fraile}, {Fraser}, {Fr{\'e}zouls}, {Gai}, {Galleti}, {Garabato},
  {Garc{\'\i}a-Sedano}, {Garofalo}, {Garralda}, {Gavel}, {Gavras}, {Gerssen},
  {Geyer}, {Giacobbe}, {Gilmore}, {Girona}, {Giuffrida}, {Glass}, {Gomes},
  {Granvik}, {Gueguen}, {Guerrier}, {Guiraud}, {Guti{\'e}rrez-S{\'a}nchez},
  {Haigron}, {Hatzidimitriou}, {Hauser}, {Haywood}, {Heiter}, {Helmi}, {Heu},
  {Hilger}, {Hobbs}, {Hofmann}, {Holland}, {Huckle}, {Hypki}, {Icardi},
  {Jan{\ss}en}, {Jevardat de Fombelle}, {Jonker}, {Juh{\'a}sz}, {Julbe},
  {Karampelas}, {Kewley}, {Klar}, {Kochoska}, {Kohley}, {Kolenberg},
  {Kontizas}, {Kontizas}, {Koposov}, {Kordopatis}, {Kostrzewa-Rutkowska},
  {Koubsky}, {Lambert}, {Lanza}, {Lasne}, {Lavigne}, {Le Fustec}, {Le
  Poncin-Lafitte}, {Lebreton}, {Leccia}, {Leclerc}, {Lecoeur-Taibi},
  {Lenhardt}, {Leroux}, {Liao}, {Licata}, {Lindstr{\o}m}, {Lister}, {Livanou},
  {Lobel}, {L{\'o}pez}, {Managau}, {Mann}, {Mantelet}, {Marchal}, {Marchant},
  {Marconi}, {Marinoni}, {Marschalk{\'o}}, {Marshall}, {Martino}, {Marton},
  {Mary}, {Massari}, {Matijevi{\v{c}}}, {Mazeh}, {McMillan}, {Messina},
  {Michalik}, {Millar}, {Molina}, {Molinaro}, {Moln{\'a}r}, {Montegriffo},
  {Mor}, {Morbidelli}, {Morel}, {Morris}, {Mulone}, {Muraveva}, {Musella},
  {Nelemans}, {Nicastro}, {Noval}, {O'Mullane}, {Ord{\'e}novic},
  {Ord{\'o}{\~n}ez-Blanco}, {Osborne}, {Pagani}, {Pagano}, {Pailler},
  {Palacin}, {Palaversa}, {Panahi}, {Pawlak}, {Piersimoni}, {Pineau}, {Plachy},
  {Plum}, {Poggio}, {Poujoulet}, {Pr{\v{s}}a}, {Pulone}, {Racero}, {Ragaini},
  {Rambaux}, {Ramos-Lerate}, {Regibo}, {Reyl{\'e}}, {Riclet}, {Ripepi}, {Riva},
  {Rivard}, {Rixon}, {Roegiers}, {Roelens}, {Romero-G{\'o}mez}, {Rowell},
  {Royer}, {Ruiz-Dern}, {Sadowski}, {Sagrist{\`a} Sell{\'e}s}, {Sahlmann},
  {Salgado}, {Salguero}, {Sanna}, {Santana-Ros}, {Sarasso}, {Savietto},
  {Schultheis}, {Sciacca}, {Segol}, {Segovia}, {S{\'e}gransan}, {Shih},
  {Siltala}, {Silva}, {Smart}, {Smith}, {Solano}, {Solitro}, {Sordo}, {Soria
  Nieto}, {Souchay}, {Spagna}, {Spoto}, {Stampa}, {Steele},
  {Steidelm{\"u}ller}, {Stephenson}, {Stoev}, {Suess}, {Surdej}, {Szabados},
  {Szegedi-Elek}, {Tapiador}, {Taris}, {Tauran}, {Taylor}, {Teixeira},
  {Terrett}, {Teyssand ier}, {Thuillot}, {Titarenko}, {Torra Clotet}, {Turon},
  {Ulla}, {Utrilla}, {Uzzi}, {Vaillant}, {Valentini}, {Valette}, {van Elteren},
  {Van Hemelryck}, {van Leeuwen}, {Vaschetto}, {Vecchiato}, {Veljanoski},
  {Viala}, {Vicente}, {Vogt}, {von Essen}, {Voss}, {Votruba}, {Voutsinas},
  {Walmsley}, {Weiler}, {Wertz}, {Wevers}, {Wyrzykowski}, {Yoldas},
  {{\v{Z}}erjal}, {Ziaeepour}, {Zorec}, {Zschocke}, {Zucker}, {Zurbach}, \&
  {Zwitter}}]{Gaia2018}
{Gaia Collaboration}, {Brown}, A.~G.~A., {Vallenari}, A., {et~al.} 2018, \aap,
  616, A1

\bibitem[{{Garufi} {et~al.}(2018){Garufi}, {Benisty}, {Pinilla}, {Tazzari},
  {Dominik}, {Ginski}, {Henning}, {Kral}, {Langlois}, {M{\'e}nard}, {Stolker},
  {Szulagyi}, {Villenave}, \& {van der Plas}}]{Garufi2018}
{Garufi}, A., {Benisty}, M., {Pinilla}, P., {et~al.} 2018, \aap, 620, A94

\bibitem[{{Garufi} {et~al.}(2017){Garufi}, {Meeus}, {Benisty}, {Quanz},
  {Banzatti}, {Kama}, {Canovas}, {Eiroa}, {Schmid}, {Stolker}, {Pohl},
  {Rigliaco}, {M{\'e}nard}, {Meyer}, {van Boekel}, \& {Dominik}}]{Garufi2017}
{Garufi}, A., {Meeus}, G., {Benisty}, M., {et~al.} 2017, \aap, 603, A21

\bibitem[{{Garufi} {et~al.}(2014){Garufi}, {Quanz}, {Schmid}, {Avenhaus},
  {Buenzli}, \& {Wolf}}]{Garufi2014b}
{Garufi}, A., {Quanz}, S.~P., {Schmid}, H.~M., {et~al.} 2014, \aap, 568, A40

\bibitem[{{Ghez} {et~al.}(1997){Ghez}, {McCarthy}, {Patience}, \&
  {Beck}}]{Ghez1997}
{Ghez}, A.~M., {McCarthy}, D.~W., {Patience}, J.~L., \& {Beck}, T.~L. 1997,
  \apj, 481, 378

\bibitem[{{Haikala} {et~al.}(2005){Haikala}, {Harju}, {Mattila}, \&
  {Toriseva}}]{Haikala2005}
{Haikala}, L.~K., {Harju}, J., {Mattila}, K., \& {Toriseva}, M. 2005, \aap,
  431, 149

\bibitem[{{Hashimoto} {et~al.}(2011){Hashimoto}, {Tamura}, {Muto}, {Kudo},
  {Fukagawa}, {Fukue}, {Goto}, {Grady}, {Henning}, {Hodapp}, {Honda},
  {Inutsuka}, {Kokubo}, {Knapp}, {McElwain}, {Momose}, {Ohashi}, {Okamoto},
  {Takami}, {Turner}, {Wisniewski}, {Janson}, {Abe}, {Brandner}, {Carson},
  {Egner}, {Feldt}, {Golota}, {Guyon}, {Hayano}, {Hayashi}, {Hayashi}, {Ishii},
  {Kandori}, {Kusakabe}, {Matsuo}, {Mayama}, {Miyama}, {Morino}, {Moro-Martin},
  {Nishimura}, {Pyo}, {Suto}, {Suzuki}, {Takato}, {Terada}, {Thalmann},
  {Tomono}, {Watanabe}, {Yamada}, {Takami}, \& {Usuda}}]{Hashimoto2011}
{Hashimoto}, J., {Tamura}, M., {Muto}, T., {et~al.} 2011, \apjl, 729, L17

\bibitem[{{Hauschildt} {et~al.}(1999){Hauschildt}, {Allard}, \&
  {Baron}}]{Hauschildt1999}
{Hauschildt}, P.~H., {Allard}, F., \& {Baron}, E. 1999, \apj, 512, 377

\bibitem[{{Henning} {et~al.}(1993){Henning}, {Pfau}, {Zinnecker}, \&
  {Prusti}}]{Henning1993}
{Henning}, T., {Pfau}, W., {Zinnecker}, H., \& {Prusti}, T. 1993, \aap, 276,
  129

\bibitem[{{Herbig}(1977)}]{Herbig1977}
{Herbig}, G.~H. 1977, \apj, 214, 747

\bibitem[{{Herczeg} \& {Hillenbrand}(2014)}]{Herczeg2014}
{Herczeg}, G.~J. \& {Hillenbrand}, L.~A. 2014, \apj, 786, 97

\bibitem[{{Hillenbrand}(1997)}]{Hillenbrand1997}
{Hillenbrand}, L.~A. 1997, \aj, 113, 1733

\bibitem[{{Jensen} {et~al.}(2009){Jensen}, {Cohen}, \&
  {Gagn{\'e}}}]{Jensen2009}
{Jensen}, E. L.~N., {Cohen}, D.~H., \& {Gagn{\'e}}, M. 2009, \apj, 703, 252

\bibitem[{{Keppler} {et~al.}(2018){Keppler}, {Benisty}, {M{\"u}ller},
  {Henning}, {van Boekel}, {Cantalloube}, {Ginski}, {van Holstein}, {Maire},
  {Pohl}, {Samland}, {Avenhaus}, {Baudino}, {Boccaletti}, {de Boer},
  {Bonnefoy}, {Chauvin}, {Desidera}, {Langlois}, {Lazzoni}, {Marleau},
  {Mordasini}, {Pawellek}, {Stolker}, {Vigan}, {Zurlo}, {Birnstiel},
  {Brandner}, {Feldt}, {Flock}, {Girard}, {Gratton}, {Hagelberg}, {Isella},
  {Janson}, {Juhasz}, {Kemmer}, {Kral}, {Lagrange}, {Launhardt}, {Matter},
  {M{\'e}nard}, {Milli}, {Molli{\`e}re}, {Olofsson}, {Perez}, {Pinilla},
  {Pinte}, {Quanz}, {Schmidt}, {Udry}, {Wahhaj}, {Williams}, {Buenzli},
  {Cudel}, {Dominik}, {Galicher}, {Kasper}, {Lannier}, {Mesa}, {Mouillet},
  {Peretti}, {Perrot}, {Salter}, {Sissa}, {Wildi}, {Abe}, {Antichi},
  {Augereau}, {Baruffolo}, {Baudoz}, {Bazzon}, {Beuzit}, {Blanchard}, {Brems},
  {Buey}, {De Caprio}, {Carbillet}, {Carle}, {Cascone}, {Cheetham}, {Claudi},
  {Costille}, {Delboulb{\'e}}, {Dohlen}, {Fantinel}, {Feautrier}, {Fusco},
  {Giro}, {Gisler}, {Gluck}, {Gry}, {Hubin}, {Hugot}, {Jaquet}, {Le Mignant},
  {Llored}, {Madec}, {Magnard}, {Martinez}, {Maurel}, {Meyer},
  {Moeller-Nilsson}, {Moulin}, {Mugnier}, {Origne}, {Pavlov}, {Perret},
  {Petit}, {Pragt}, {Puget}, {Rabou}, {Ramos}, {Rigal}, {Rochat}, {Roelfsema},
  {Rousset}, {Roux}, {Salasnich}, {Sauvage}, {Sevin}, {Soenke}, {Stadler},
  {Suarez}, {Turatto}, \& {Weber}}]{Keppler2018}
{Keppler}, M., {Benisty}, M., {M{\"u}ller}, A., {et~al.} 2018, ArXiv e-prints
  [\eprint[arXiv]{1806.11568}]

\bibitem[{{K{\"o}hler} {et~al.}(2000){K{\"o}hler}, {Kunkel}, {Leinert}, \&
  {Zinnecker}}]{Koehler2000}
{K{\"o}hler}, R., {Kunkel}, M., {Leinert}, C., \& {Zinnecker}, H. 2000, \aap,
  356, 541

\bibitem[{{Kurtovic} {et~al.}(2018){Kurtovic}, {P{\'e}rez}, {Benisty}, {Zhu},
  {Zhang}, {Huang}, {Andrews}, {Dullemond}, {Isella}, {Bai}, {Carpenter},
  {Guzm{\'a}n}, {Ricci}, \& {Wilner}}]{Kurtovic2018}
{Kurtovic}, N.~T., {P{\'e}rez}, L.~M., {Benisty}, M., {et~al.} 2018, \apjl,
  869, L44

\bibitem[{{Langlois} {et~al.}(2014){Langlois}, {Dohlen}, {Vigan}, {Zurlo},
  {Moutou}, {Schmid}, {Mili}, {Beuzit}, {Boccaletti}, {Carle}, {Costille},
  {Dorn}, {Gluck}, {Hubin}, {Feldt}, {Kasper}, {Lizon}, {Madec}, {Le Mignant},
  {Mouillet}, {Puget}, {Sauvage}, \& {Wildi}}]{Langlois2014}
{Langlois}, M., {Dohlen}, K., {Vigan}, A., {et~al.} 2014, in Society of
  Photo-Optical Instrumentation Engineers (SPIE) Conference Series, Vol. 9147,
  \procspie, 91471R

\bibitem[{{Langlois} {et~al.}(2018){Langlois}, {Pohl}, {Lagrange}, {Maire},
  {Mesa}, {Boccaletti}, {Gratton}, {Denneulin}, {Klahr}, {Vigan}, {Benisty},
  {Dominik}, {Bonnefoy}, {Menard}, {Avenhaus}, {Cheetham}, {Van Boekel}, {de
  Boer}, {Chauvin}, {Desidera}, {Feldt}, {Galicher}, {Ginski}, {Girard},
  {Henning}, {Janson}, {Kopytova}, {Kral}, {Ligi}, {Messina}, {Peretti},
  {Pinte}, {Sissa}, {Stolker}, {Zurlo}, {Magnard}, {Blanchard}, {Buey},
  {Suarez}, {Cascone}, {Moller-Nilsson}, {Weber}, {Petit}, \&
  {Pragt}}]{Langlois2018}
{Langlois}, M., {Pohl}, A., {Lagrange}, A.-M., {et~al.} 2018, ArXiv e-prints
  [\eprint[arXiv]{1802.03995}]

\bibitem[{{Loinard} {et~al.}(2008){Loinard}, {Torres}, {Mioduszewski}, \&
  {Rodr{\'\i}guez}}]{Loinard2008}
{Loinard}, L., {Torres}, R.~M., {Mioduszewski}, A.~J., \& {Rodr{\'\i}guez},
  L.~F. 2008, \apjl, 675, L29

\bibitem[{{Long} {et~al.}(2018){Long}, {Herczeg}, {Pascucci}, {Apai},
  {Henning}, {Manara}, {Mulders}, {Sz{\H u}cs}, \& {Hendler}}]{Long2018}
{Long}, F., {Herczeg}, G.~J., {Pascucci}, I., {et~al.} 2018, ArXiv e-prints
  [\eprint[arXiv]{1806.04826}]

\bibitem[{{Luhman}(2004)}]{Luhman2004}
{Luhman}, K.~L. 2004, \apj, 602, 816

\bibitem[{{Luhman}(2007)}]{Luhman2007}
{Luhman}, K.~L. 2007, \apjs, 173, 104

\bibitem[{{Maheswar} {et~al.}(2003){Maheswar}, {Manoj}, \&
  {Bhatt}}]{Maheswar2003}
{Maheswar}, G., {Manoj}, P., \& {Bhatt}, H.~C. 2003, \aap, 402, 963

\bibitem[{{Maire} {et~al.}(2016){Maire}, {Bonnefoy}, {Ginski}, {Vigan},
  {Messina}, {Mesa}, {Galicher}, {Gratton}, {Desidera}, {Kopytova}, {Millward},
  {Thalmann}, {Claudi}, {Ehrenreich}, {Zurlo}, {Chauvin}, {Antichi},
  {Baruffolo}, {Bazzon}, {Beuzit}, {Blanchard}, {Boccaletti}, {de Boer},
  {Carle}, {Cascone}, {Costille}, {De Caprio}, {Delboulb{\'e}}, {Dohlen},
  {Dominik}, {Feldt}, {Fusco}, {Girard}, {Giro}, {Gisler}, {Gluck}, {Gry},
  {Henning}, {Hubin}, {Hugot}, {Jaquet}, {Kasper}, {Lagrange}, {Langlois}, {Le
  Mignant}, {Llored}, {Madec}, {Martinez}, {Mawet}, {Milli},
  {M{\"o}ller-Nilsson}, {Mouillet}, {Moulin}, {Moutou}, {Orign{\'e}}, {Pavlov},
  {Petit}, {Pragt}, {Puget}, {Ramos}, {Rochat}, {Roelfsema}, {Salasnich},
  {Sauvage}, {Schmid}, {Turatto}, {Udry}, {Vakili}, {Wahhaj}, {Weber}, \&
  {Wildi}}]{Maire2016}
{Maire}, A.~L., {Bonnefoy}, M., {Ginski}, C., {et~al.} 2016, \aap, 587, A56

\bibitem[{{Manara} {et~al.}(2019){Manara}, {Tazzari}, {Long}, {Herczeg},
  {Lodato}, {Rota}, {Cazzoletti}, {van der Plas}, {Pinilla}, {Dipierro},
  {Edwards}, {Harsono}, {Johnstone}, {Liu}, {Menard}, {Nisini}, {Ragusa},
  {Boehler}, \& {Cabrit}}]{Manara2019}
{Manara}, C.~F., {Tazzari}, M., {Long}, F., {et~al.} 2019, \aap, 628, A95

\bibitem[{{Marino} {et~al.}(2015){Marino}, {Perez}, \& {Casassus}}]{Marino2015}
{Marino}, S., {Perez}, S., \& {Casassus}, S. 2015, \apjl, 798, L44

\bibitem[{{Mayama} {et~al.}(2012){Mayama}, {Hashimoto}, {Muto}, {Tsukagoshi},
  {Kusakabe}, {Kuzuhara}, {Takahashi}, {Kudo}, {Dong}, {Fukagawa}, {Takami},
  {Momose}, {Wisniewski}, {Follette}, {Abe}, {Akiyama}, {Brandner}, {Brandt},
  {Carson}, {Egner}, {Feldt}, {Goto}, {Grady}, {Guyon}, {Hayano}, {Hayashi},
  {Hayashi}, {Henning}, {Hodapp}, {Ishii}, {Iye}, {Janson}, {Kandori}, {Kwon},
  {Knapp}, {Matsuo}, {McElwain}, {Miyama}, {Morino}, {Moro-Martin},
  {Nishimura}, {Pyo}, {Serabyn}, {Suto}, {Suzuki}, {Takato}, {Terada},
  {Thalmann}, {Tomono}, {Turner}, {Watanabe}, {Yamada}, {Takami}, {Usuda}, \&
  {Tamura}}]{Mayama2012}
{Mayama}, S., {Hashimoto}, J., {Muto}, T., {et~al.} 2012, \apjl, 760, L26

\bibitem[{{Metchev} \& {Hillenbrand}(2009)}]{Metchev2009}
{Metchev}, S.~A. \& {Hillenbrand}, L.~A. 2009, \apjs, 181, 62

\bibitem[{{Murakawa}(2010)}]{Murakawa2010}
{Murakawa}, K. 2010, \aap, 518, A63

\bibitem[{{Muro-Arena} {et~al.}(2018){Muro-Arena}, {Dominik}, {Waters}, {Min},
  {Klarmann}, {Ginski}, {Isella}, {Benisty}, {Pohl}, {Garufi}, {Hagelberg},
  {Langlois}, {Menard}, {Pinte}, {Sezestre}, {van der Plas}, {Villenave},
  {Delboulb{\'e}}, {Magnard}, {M{\"o}ller-Nilsson}, {Pragt}, {Rabou}, \&
  {Roelfsema}}]{MuroArena2018}
{Muro-Arena}, G.~A., {Dominik}, C., {Waters}, L.~B.~F.~M., {et~al.} 2018, \aap,
  614, A24

\bibitem[{{Muto} {et~al.}(2012){Muto}, {Grady}, {Hashimoto}, {Fukagawa},
  {Hornbeck}, {Sitko}, {Russell}, {Werren}, {Cur{\'e}}, {Currie}, {Ohashi},
  {Okamoto}, {Momose}, {Honda}, {Inutsuka}, {Takeuchi}, {Dong}, {Abe},
  {Brandner}, {Brandt}, {Carson}, {Egner}, {Feldt}, {Fukue}, {Goto}, {Guyon},
  {Hayano}, {Hayashi}, {Hayashi}, {Henning}, {Hodapp}, {Ishii}, {Iye},
  {Janson}, {Kandori}, {Knapp}, {Kudo}, {Kusakabe}, {Kuzuhara}, {Matsuo},
  {Mayama}, {McElwain}, {Miyama}, {Morino}, {Moro-Martin}, {Nishimura}, {Pyo},
  {Serabyn}, {Suto}, {Suzuki}, {Takami}, {Takato}, {Terada}, {Thalmann},
  {Tomono}, {Turner}, {Watanabe}, {Wisniewski}, {Yamada}, {Takami}, {Usuda}, \&
  {Tamura}}]{Muto2012}
{Muto}, T., {Grady}, C.~A., {Hashimoto}, J., {et~al.} 2012, \apjl, 748, L22

\bibitem[{{Pascucci} {et~al.}(2016){Pascucci}, {Testi}, {Herczeg}, {Long},
  {Manara}, {Hendler}, {Mulders}, {Krijt}, {Ciesla}, {Henning}, {Mohanty},
  {Drabek-Maunder}, {Apai}, {Sz{\H u}cs}, {Sacco}, \&
  {Olofsson}}]{Pascucci2016}
{Pascucci}, I., {Testi}, L., {Herczeg}, G.~J., {et~al.} 2016, \apj, 831, 125

\bibitem[{{P{\'e}rez} {et~al.}(2016){P{\'e}rez}, {Carpenter}, {Andrews},
  {Ricci}, {Isella}, {Linz}, {Sargent}, {Wilner}, {Henning}, {Deller},
  {Chandler}, {Dullemond}, {Lazio}, {Menten}, {Corder}, {Storm}, {Testi},
  {Tazzari}, {Kwon}, {Calvet}, {Greaves}, {Harris}, \& {Mundy}}]{Perez2016}
{P{\'e}rez}, L.~M., {Carpenter}, J.~M., {Andrews}, S.~M., {et~al.} 2016,
  Science, 353, 1519

\bibitem[{{Quanz} {et~al.}(2013){Quanz}, {Avenhaus}, {Buenzli}, {Garufi},
  {Schmid}, \& {Wolf}}]{Quanz2013b}
{Quanz}, S.~P., {Avenhaus}, H., {Buenzli}, E., {et~al.} 2013, \apjl, 766, L2

\bibitem[{{Rapson} {et~al.}(2015){Rapson}, {Kastner}, {Andrews}, {Hines},
  {Macintosh}, {Millar-Blanchaer}, \& {Tamura}}]{Rapson2015}
{Rapson}, V.~A., {Kastner}, J.~H., {Andrews}, S.~M., {et~al.} 2015, \apjl, 803,
  L10

\bibitem[{{Ratzka} {et~al.}(2005){Ratzka}, {K{\"o}hler}, \&
  {Leinert}}]{Ratzka2005}
{Ratzka}, T., {K{\"o}hler}, R., \& {Leinert}, C. 2005, \aap, 437, 611

\bibitem[{{Ricci} {et~al.}(2010){Ricci}, {Testi}, {Natta}, \&
  {Brooks}}]{Ricci2010}
{Ricci}, L., {Testi}, L., {Natta}, A., \& {Brooks}, K.~J. 2010, \aap, 521, A66

\bibitem[{{Stolker} {et~al.}(2016{\natexlab{a}}){Stolker}, {Dominik},
  {Avenhaus}, {Min}, {de Boer}, {Ginski}, {Schmid}, {Juhasz}, {Bazzon},
  {Waters}, {Garufi}, {Augereau}, {Benisty}, {Boccaletti}, {Henning},
  {Langlois}, {Maire}, {M{\'e}nard}, {Meyer}, {Pinte}, {Quanz}, {Thalmann},
  {Beuzit}, {Carbillet}, {Costille}, {Dohlen}, {Feldt}, {Gisler}, {Mouillet},
  {Pavlov}, {Perret}, {Petit}, {Pragt}, {Rochat}, {Roelfsema}, {Salasnich},
  {Soenke}, \& {Wildi}}]{Stolker2016a}
{Stolker}, T., {Dominik}, C., {Avenhaus}, H., {et~al.} 2016{\natexlab{a}},
  \aap, 595, A113

\bibitem[{{Stolker} {et~al.}(2016{\natexlab{b}}){Stolker}, {Dominik}, {Min},
  {Garufi}, {Mulders}, \& {Avenhaus}}]{Stolker2016b}
{Stolker}, T., {Dominik}, C., {Min}, M., {et~al.} 2016{\natexlab{b}}, \aap,
  596, A70

\bibitem[{{Torres} {et~al.}(2006){Torres}, {Quast}, {da Silva}, {de La Reza},
  {Melo}, \& {Sterzik}}]{Torres2006}
{Torres}, C.~A.~O., {Quast}, G.~R., {da Silva}, L., {et~al.} 2006, \aap, 460,
  695

\bibitem[{{van Boekel} {et~al.}(2017){van Boekel}, {Henning}, {Menu}, {de
  Boer}, {Langlois}, {M{\"u}ller}, {Avenhaus}, {Boccaletti}, {Schmid},
  {Thalmann}, {Benisty}, {Dominik}, {Ginski}, {Girard}, {Gisler}, {Lobo Gomes},
  {Menard}, {Min}, {Pavlov}, {Pohl}, {Quanz}, {Rabou}, {Roelfsema}, {Sauvage},
  {Teague}, {Wildi}, \& {Zurlo}}]{vanBoekel2017}
{van Boekel}, R., {Henning}, T., {Menu}, J., {et~al.} 2017, \apj, 837, 132

\bibitem[{{van der Plas} {et~al.}(2017){van der Plas}, {Wright}, {M{\'e}nard},
  {Casassus}, {Canovas}, {Pinte}, {Maddison}, {Maaskant}, {Avenhaus}, {Cieza},
  {Perez}, \& {Ubach}}]{vanderPlas2016}
{van der Plas}, G., {Wright}, C.~M., {M{\'e}nard}, F., {et~al.} 2017, \aap,
  597, A32

\bibitem[{{van Holstein} {et~al.}(2019){van Holstein}, {Girard}, {de Boer},
  {Snik}, {Milli}, {Stam}, {Ginski}, {Mouillet}, {Wahhaj}, {Schmid}, {Keller},
  {Langlois}, {Dohlen}, {Vigan}, {Pohl}, {Carbillet}, {Fantinel}, {Maurel},
  {Orign{\'e}}, {Petit}, {Ramos}, {Rigal}, {Sevin}, {Boccaletti}, {Le
  Coroller}, {Dominik}, {Henning}, {Lagadec}, {M{\'e}nard}, {Turatto}, {Udry},
  {Chauvin}, {Feldt}, \& {Beuzit}}]{vanHolstein2019}
{van Holstein}, R.~G., {Girard}, J.~H., {de Boer}, J., {et~al.} 2019, arXiv
  e-prints, arXiv:1909.13108

\bibitem[{{van Terwisga} {et~al.}(2018){van Terwisga}, {van Dishoeck},
  {Ansdell}, {van der Marel}, {Testi}, {Williams}, {Facchini}, {Tazzari},
  {Hogerheijde}, {Trapman}, {Manara}, {Miotello}, {Maud}, \&
  {Harsono}}]{vanTerwisga2018}
{van Terwisga}, S.~E., {van Dishoeck}, E.~F., {Ansdell}, M., {et~al.} 2018,
  \aap, 616, A88

\end{thebibliography}

\begin{appendix}

\section{Stellar companions} \label{Appendix_companion}
Thirteen of our targets show at least one visual companion in our intensity images. In total, we found 28 companions. However, 18 of these are very faint (i.e., lower than 1\% in flux with respect to the primary) and are not considered in this work because they are likely background objects. The other 10 objects are physical companions of 9 primary stars, as shown in Table \ref{Sample_properties}. The astrometry of all the visual companions in our images is listed in Table \ref{Companion_table}. 

The flux of the companion relative to the primary is calculated from the non-coronagraphic flux frames, where the two fluxes are in direct relation. Faint sources are not visible in these frames, and we were only able to set the upper limit of 1\% from the science frames where the flux of the primary star is indeed a lower limit because of the coronagraph. The relative flux of the bright close companions was used to calculate the stellar luminosity, as described in Sect.\,\ref{Stellar_properties}. In one case (IK Lup), the flux and precise astrometry of a bright companion at $\sim$6.3\arcsec\ could not be extracted because the star lies at the edge of our detector. However, this object is likely bound to IK Lup because \textit{Gaia} detected it and provided a parallax comparable to that of the primary star (6.36 vs. 6.44). In addition to these sources, DoAr 21 is a possible spectroscopic binary \citep{Loinard2008} that is obviously not resolved by our images.

\begin{table}
\centering
\caption{List of visual companions in our images. Columns are the primary star, the separation, the P.A., and the flux of the companion relative to the primary. The second item lies at the detector edge, and we were unable to calculate the flux.} 
\label{Companion_table}
\begin{tabular}{cccc}
\hline
Target & $r$ (\arcsec) & P.A.\,($\degree$) & $F_2/F_1$ \\
\hline
IK Lup & 5.48 & 171.0 & <0.01 \\
IK Lup & $\sim$6.3 & $\sim$97 & - \\
HT Lup & 0.16 & 247.0 & 0.18 \\
HT Lup & 2.80 & 296.9 & 0.07 \\
J1606-1928 & 0.57 & 135.8 & 0.77 \\
J1606-1908 & 0.21 & 210.8 & 0.14 \\
HK Lup & 2.43 & 206.6 & <0.01 \\
HK Lup & 2.74 & 264.7 & <0.01 \\
HK Lup & 3.06 & 36.1 & <0.01 \\
HK Lup & 3.85 & 277.5 & <0.01 \\
HK Lup & 3.91 & 312.4 & <0.01 \\
HK Lup & 4.14 & 315.1 & <0.01 \\
HK Lup & 5.25 & 42.5 & <0.01 \\
V1094 Sco & 4.21 & 224.5 & <0.01 \\
V1094 Sco & 5.42 & 317.6 & <0.01 \\
J1610-1904 & 0.26 & 58.9 & 0.64 \\
J1610-1904 & 6.73 & 232.3 & <0.01 \\
J1614-2305 & 0.35 & 109.8 & 1.1 \\
J1614-2305 & 2.92 & 98.2 & <0.01 \\
VV Sco & 1.87 & 338.4 & 0.37 \\
VV Sco & 6.13 & 223.1 & <0.01 \\
J1615-1921 & 2.87 & 175.5 & <0.01 \\
J1615-1921 & 5.53 & 74.4 & <0.01 \\
DoAr16 & 0.81 & 35.4 & 0.41 \\
DoAr16 & 1.45 & 122.4 & <0.01 \\
SR4 & 5.60 & 225.0 & <0.01 \\
DoAr25 & 4.43 & 325.5 & <0.01 \\
SR9 & 0.66 & 356.4 & 0.08 \\
\hline
\end{tabular} 
\end{table}

\section{Observing setup}
Table \ref{Observing_setup} shows the observing setup of our sample.

\section{Contrast of polarized to stellar light} \label{Appendix_contrast}
We used the contrast of polarized to stellar light as described in detail by \citet{Garufi2017}. This number is useful for evaluating the fraction of polarized scattered photons from a certain disk location over the total photons incident on that disk region. Because the stellar light becomes diluted with distance, this term is described by $F_{\rm pol}\cdot 4\pi r^2/F_*$. This number is averaged over the radii between the inner inset of the disk (in case of a disk cavity, otherwise the coronagraphic size) and the location with the outermost detectable signal. To mitigate the effect of the scattering phase function on this number, this calculation is performed along the disk major axis where photons are always scattered by an angle close to $90\degree$. In this way, more strongly inclined disks are not artificially brighter than face-on disks because of photons scattered by smaller angles (of which there are many more because scattering phase functions are forward peaking). 

In case of coronagraphic images (as are those of this work), a further complication is the calculation of the stellar luminosity in the same detector units as the $Q_{\phi}$ images. This problem is overcome by calculating the luminosity from the complementary flux frames that are taken together with the science frames. This number is then converted into the proper luminosity considering the different exposure times of flux and science frames as well as the possible presence of a neutral density filter that is often used when flux frames are taken. The errors tabulated in Table \ref{Sample_properties} do not account for the uncertainty deriving from this procedure.

\begin{table*}
\centering
\caption{Summary of observations. Columns are the reference number in this work, the target name, the observation date, the individual integration time (DIT) multiplied by the number of integrations (NDIT) and the number of polarimetric cycles (NCYCLES), the total integration time, and DIMM seeing during the observations. The total integration time is obtained from DIT (s) $\times$ NDIT $\times$ NCYCLE $\times$ 4 because each polarimetric cycle is composed of four Stokes setups ($Q+$, $Q-$, $U+$, and $U-$). } 
\label{Observing_setup}
\begin{tabular}{ccccccc}
\hline
Reference & Target & Observation & DIT (s) $\times$ NDIT $\times$ NCYCLE & Exposure & DIMM seeing \\
number & name & date & & time (min) & (\arcsec) \\
\hline
1 & WW Cha & 12-03-17 & 32 $\times$ 2 $\times$ 6 & 25.6 & 0.42$-$0.55 \\
2 & Sz 45 & 13-03-17 & 96 $\times$ 1 $\times$ 6 & 38.4 & 0.45$-$0.78 \\
3 & IK Lup & 07-03-17 & 96 $\times$ 1 $\times$ 4 & 25.6 & 0.36$-$0.50 \\
4 & HT Lup & 13-03-17 & 64 $\times$ 1 $\times$ 9 & 38.4 & 0.39$-$0.73 \\
5 & 2MASS J16035767-2031055 & 11-03-17 & 96 $\times$ 1 $\times$ 4 & 25.6 & 0.33$-$0.49 \\
6 & 2MASS J16062196-1928445 & 13-03-17 & 96 $\times$ 1 $\times$ 5 & 32.0 & 0.41$-$0.67 \\
7 & 2MASS J16064385-1908056 & 07-03-17 & 96 $\times$ 1 $\times$ 4 & 25.6 & 0.40$-$0.50 \\
8 & HK Lup & 11-03-17 & 96 $\times$ 1 $\times$ 6 & 38.4 & 0.34$-$0.46 \\
9 & V1094 Sco & 12-03-17 & 96 $\times$ 1 $\times$ 6 & 38.4 & 0.36$-$0.51 \\
10 & 2MASS J16090075-1908526 & 07-03-17 & 96 $\times$ 1 $\times$ 4 & 25.6 & 0.47$-$0.62 \\
11 & 2MASS J16102857-1904469 & 07-03-17 & 32 $\times$ 2 $\times$ 6 & 25.6 & 0.63$-$0.94 \\
12 & 2MASS J16111534-1757214 & 07-03-17 & 96 $\times$ 1 $\times$ 6 & 38.4 & 0.40$-$0.63 \\
13 & 2MASS J16141107-2305362 & 13-03-17 & 64 $\times$ 2 $\times$ 4 & 34.1 & 0.43$-$0.58 \\
\multirow{2}{*}{14} & \multirow{2}{*}{2MASS J16142029-1906481} & \multirow{2}{*}{13-03-17} & 64 $\times$ 1 $\times$ 5 & \multirow{2}{*}{34.1} & \multirow{2}{*}{0.50$-$0.60} \\
 & & & 96 $\times$ 1 $\times$ 2 & & & \\ 
15 & VV Sco & 12-03-17 & 64 $\times$ 1 $\times$ 8 & 34.1 & 0.45$-$0.54 \\
16 & 2MASS J16154416-1921171 & 12-03-17 & 96 $\times$ 1 $\times$ 6 & 38.4 & 0.44$-$0.58 \\
17 & DoAr 16 & 11-03-17 & 64 $\times$ 1 $\times$ 6 & 25.6 & 0.54$-$0.91 \\
18 & EM* SR4 & 12-03-17 & 32 $\times$ 2 $\times$ 8 & 34.1 & 0.40$-$0.55 \\
\multirow{2}{*}{19} & \multirow{2}{*}{DoAr 21} & \multirow{2}{*}{11-03-17} & 16 $\times$ 4 $\times$ 2 & \multirow{2}{*}{34.1} & \multirow{2}{*}{0.38$-$0.58} \\ 
& & & 32 $\times$ 2 $\times$ 6 & &  \\ 
20 & DoAr 25 & 07-03-17 & 96 $\times$ 1 $\times$ 6 & 38.4 & 0.50$-$0.67 \\
21 & EM* SR9 & 11-03-17 & 96 $\times$ 1 $\times$ 6 & 38.4 & 0.37$-$0.73 \\
\hline
\end{tabular} 
\end{table*}

\end{appendix}

\end{document}